\documentclass[a4paper,11pt,DIV14,numbers=noendperiod]{scrartcl} 
\pdfoutput=1

\usepackage{amsmath,amssymb,amsfonts}
\usepackage{epsfig,cancel,cite}
\usepackage[usenames,dvipsnames]{color}
\usepackage{hyperref} 

\usepackage{dsfont} 
\usepackage{bbm} 

\usepackage{amsmath}
\usepackage{amsfonts}
\usepackage{amssymb}
\usepackage{amsthm}

\usepackage{graphicx}
\usepackage{booktabs}
\usepackage{epstopdf}
\newlength{\xtrawidth}
\newlength{\xtraheight}
\newcommand{\quot}[1]{``#1''}

\numberwithin{equation}{section}

\usepackage{slashed} 
\def\d{\textrm{d}}
\newcommand{\DD}{\mathrm{D}}
\newcommand{\I}{\mathrm{i}}
\def\tr{\textrm{tr}\,}

\newcommand{\SO}[1]{\ensuremath{\mathrm{SO}(#1)}}
\newcommand{\SU}[1]{\ensuremath{\mathrm{SU}(#1)}}
\newcommand{\U}[1]{\ensuremath{\mathrm{U}(#1)}}
\newcommand{\Sp}[1]{\ensuremath{\mathrm{Sp}(#1)}}
\newcommand{\Gtwo}{\ensuremath{\mathrm{G}_{2}}}
\newcommand{\vol}{\mathcal{V}_0}

\begin{document}

\title{\vskip 1.5cm {\LARGE Moduli Stabilising in Heterotic Nearly K\"ahler Compactifications\\[0.5cm]}}
\author{
\bf\large Michael Klaput\footnote{Michael.Klaput@physics.ox.ac.uk},\setcounter{footnote}{3}
Andre Lukas\footnote{lukas@physics.ox.ac.uk},
Cyril Matti\footnote{cyril.matti@a3.epfl.ch} and
Eirik E.~Svanes\footnote{Eirik.Svanes@physics.ox.ac.uk}}
\date{}
\maketitle
\begin{center}{\small {\it \vskip -0.8cm Rudolf Peierls Center for Theoretical Physics, Oxford University\\}
{\it 1 Keble Road, Oxford, OX1 3NP, U.K.\\[1cm]}}
\end{center}

\thispagestyle{empty}
\setcounter{page}{0}
\abstract{\noindent
We study heterotic string compactifications on nearly K\"ahler homogeneous spaces, including the gauge field effects which arise at order $\alpha'$. Using Abelian gauge fields, we are able to solve the Bianchi identity and supersymmetry conditions to this order. The four-dimensional external space-time consists of a domain wall solution with moduli fields varying along the transverse direction. We find that the inclusion of $\alpha'$ corrections improves the moduli stabilization features of this solution. In this case, one of the dilaton and the volume modulus asymptotes to a constant value away from the domain wall. It is further shown that the inclusion of non-perturbative effects can stabilize the remaining modulus and ``lift" the domain wall to an AdS vacuum. The coset $\SU{3}/\U{1}^2$ is used as an explicit example to demonstrate the validity of this AdS vacuum. Our results show that heterotic nearly K\"ahler compactifications can lead to maximally symmetric four-dimensional space-times at the non-perturbative level.
}

\newpage

\tableofcontents

\section{Introduction}

In the search for realistic models of particle physics, the $E_8\times E_8$ heterotic string~\cite{Green1984117}, compactified on Calabi-Yau manifolds, has long been an attractive approach to string model building~\cite{Candelas198546}, due to its appealing properties for gauge coupling unification~\cite{PhysRevLett.32.438} and the ``built-in" exceptional gauge groups, among others. Indeed, large numbers of heterotic standard models have recently been constructed by compactifying on Calabi-Yau manifolds with Abelian gauge bundles~\cite{Anderson:2011ns , Anderson:2012yf}.

Moduli stabilization of heterotic Calabi-Yau compactifications has been more difficult as compared to the type IIB string (although see~\cite{Anderson:2009sw,Anderson:2009nt,Anderson:2010mh, Anderson:2011cza} for some recent progress), mainly due to the absence of RR fluxes. It is expected that the RR fluxes are mapped into geometric fluxes (torsion) on the heterotic side, motivating the study of heterotic compactifications on non Calabi-Yau spaces. The first such class of heterotic compactifications, based on complex non-K\"ahler manifolds has been studied by Strominger~\cite{Strominger1986253} and was developed further in Refs.~\cite{LopesCardoso:2002hd,Gauntlett:2003cy,Becker:2003yv,LopesCardoso:2003af,Becker:2003sh,2009CMaPh.288..677F,Martelli:2010jx}.

A more general class of heterotic non Calabi-Yau compactifications on half-flat mirror manifolds -- manifolds which appear in the context of type II mirror symmetry with NS flux -- has been introduced some time ago in Refs.~\cite{Gurrieri:2004dt, Gurrieri:2007jg , Lukas:2010mf}. Although few half-flat mirror manifolds are known explicitly some of their properties can be inferred from mirror symmetry, to the extent that compactification and computation of the associated effective theories can be carried out reasonably explicitly, a feature which is non-trivial for non Calabi-Yau compactifications. This considerable advantage comes at a price of introducing two major complications. First, half-flat mirror manifolds do not solve the heterotic string in the presence of four-dimensional maximally symmetric space-time but rather have to be combined with a four-dimensional half BPS domain wall for a full 10-dimensional solution. Secondly, it is not clear in general how to construct the gauge bundles required for heterotic compactifications on half-flat mirror manifolds. Let us discuss these two issues in turn.

At first sight a four-dimensional non maximally symmetric space-time such as a domain wall appears to be phenomenologically unviable. However, it was shown in Refs.~\cite{Gurrieri:2004dt,Gurrieri:2007jg} that heterotic compactifications on half-flat mirror manifolds can still be associated with a fully covariant four-dimensional $N=1$ supergravity theory. Due to a superpotential and an associated runaway direction present in this theory it is not solved by Minkowski or AdS space but, in the simplest case, by a domain wall which forms the four-dimensional part of the aforementioned 10-dimensional solution. Obtaining a maximally symmetric four-dimensional space-time therefore becomes a matter of lifting a runaway direction in the scalar potential of the theory by additional contributions, for example of non-perturbative origin, a task frequently required in string compactifications. In conclusion, heterotic half-flat compactifications are still potentially viable subject to such a ``lifting" being carried out successfully.

The problem of constructing gauge bundles is technical rather than conceptual in nature. Progress in this direction has been made in Refs.~\cite{Lust:1985be,Dieter1986220,1126-6708-2009-09-077,Chatzistavrakidis:2008ii,Lechtenfeld:2010dr,Klaput:2011mz} by focusing on nearly K\"ahler manifolds which are given as six-dimensional group or group coset manifolds. The most relevant example for our purpose is the coset $SU(3)/U(1)^2$. The underlying group structure of these manifolds allows for an explicit construction of certain bundles, notably line bundles, and their associated connections. In particular, it has been shown that the coset $SU(3)/U(1)^2$ with vector bundles constructed as sums of line bundles can lead to models with GUT-type symmetries and three chiral families. 

This discussion suggests two important questions which have been left unanswered in the work on heterotic half-flat compactifications to date. Can the runaway direction indeed be lifted and a maximally symmetric four-dimensional space-time be achieved? Can we understand the back-reaction of the gauge fields, induced by the Bianchi identity at order $\alpha'$, onto the other fields?

These are the two main questions which we will address in the present paper, working within the context of nearly K\"ahler spaces and, in particular, the coset space $SU(3)/U(1)^2$. We will see that the answers are ``yes" in both cases and that the two issues of moduli stabilisation and $\alpha'$ corrections are indeed related. For line bundle sums we are able to solve the Bianchi identity and compute the effect of the resulting non-vanishing NS field strength $H$ at order $\alpha'$. We find that these $\alpha'$ effects help with moduli stabilisation in that they lead to one of the relevant moduli (either the dilaton or the volume modulus) being asymptotically constant away from the domain wall. Adding non-perturbative effects from gaugino condensation then leads to a complete stabilisation and a four-dimensional AdS vacuum. For appropriate choices of the parameters (in particular the gauge bundle fluxes) we find that the internal volume is sufficiently large for the $\alpha'$ expansion to be justified. 

The plan of the paper is as follows.  In Section \ref{chap:two} we start by reviewing half-flat domain wall solutions of the heterotic string. 
Section \ref{chap:three} describes the specific form of these solutions for coset spaces at lowest order in $\alpha'$ and, in Section~\ref{chap:four}, these results are extended to first order in $\alpha'$. In Section~\ref{chap:five} we introduce the associated four-dimensional theories and discuss moduli stabilization. We conclude in Section~\ref{chap:six}. Conventions and details of the underlying calculations are provided in a number of technical appendices. 

\section{Heterotic supergravity and Hitchin flow}\label{chap:two}
Before we describe the explicit solutions to order $\alpha'$ central to this paper, we briefly discuss the general setting of $N=1$ heterotic supergravity and domain wall solutions thereof. Half-flat manifolds and, in particular, the nearly K\"ahler manifolds that we shall be concerned with later, form solutions to the heterotic equations at leading order in $\alpha'$ provided they are combined with a four-dimensional domain-wall solution~\cite{Lukas:2010mf, Klaput:2011mz}. In this case, the variation of the half-flat manifold along the direction transverse to the domain wall is described by Hitchin flow equations, as we will review. 

\subsection{Heterotic supergravity}\label{section:hetsugra}
The low-energy limit of heterotic $E_8\times E_8$ string theory is given by a 10-dimensional $N=1$ supergravity theory coupled to 10-dimensional super Yang-Mills theory with $E_8\times E_8$ gauge group. Its bosonic field content consists of the metric $g$, the dilaton $\phi$, the two-form $B$ and a $E_8\times E_8$ gauge field $A$. The corresponding action can be obtained from sigma model perturbation theory up to two loops\footnote{See also Ref.~\cite{Bergshoeff:1989de} which uses the supersymmetrisation of the Yang-Mills and Chern-Simons forms. A modern version of this derivation has been recently given in Ref.~\cite{Andriot:2011iw}.} \cite{Hull:1987pc} and its bosonic part in the string frame is given by
\begin{align}
S=-\frac{1}{2\kappa_{10}^2}\int e^{-2\phi}\Big[\mathcal{R}*1-4\d\phi\wedge*\d\phi+\frac{1}{2}H\wedge*H+\frac{\alpha'}{4}\,(\tr F\wedge*F-\tr R^{-}\wedge * R^{-})\Big]+ \mathcal{O}(\alpha'^2)\;.
\label{eq:action}
\end{align}
Here $\kappa_{10}$ is the ten-dimensional Planck constant, $F=\d A+A\wedge A$ is the gauge field strengths, $\mathcal{R}$ it the curvature scalar associated to the Levi-Civita connection $\omega$ and $R^{-}$ is the curvature two-form obtained from the connection
\begin{align}\label{eq:hullconnectiondef}
\omega_{IJ}^{-\;\;K}=\omega_{IJ}^{\;\;\;\;K}-\frac{1}{2}H_{IJ}^{\;\;\;\;K}\; ,
\end{align}
also known as \emph{Hull connection} in the literature.
\\\\
The three-form field strength $H$ is defined as
\begin{equation}
 H=\d B+\frac{\alpha'}{4}\left(w_{\rm YM}-w_{\rm Gr}\right)
\end{equation} 
with the Yang-Mills and gravity Chern-Simons forms satisfying $\d w_{\rm YM}=\tr F\wedge F$ and $\d w_{\rm Gr}=\tr R^-\wedge R^-$, respectively. Taking the exterior derivative then leads to the Bianchi identity
\begin{align}
\d H=\frac{\alpha'}{4}(\tr F\wedge F-\tr R^{-}\wedge R^{-})\;.
\label{eq:bianchi}
\end{align}

The fermionic field content of the supergravity is given by the gravitino $\psi_M$, the dilatino $\lambda$ and the gaugino $\chi$. The corresponding supersymmetry transformations are
\begin{align}
\label{eq:SUSY1zero}
\delta\psi_M
&=
\left(
\nabla_M+\frac{1}{8}\mathcal{H}_M 
\right)\,\varepsilon 
+
\mathcal{O}(\alpha'^2)
\\
\delta\lambda
&=
\left(\slashed\partial\phi+\frac{1}{12}\mathcal{H} 
\right)\,\varepsilon 
+
\mathcal{O}(\alpha'^2)
\\
\label{eq:SUSY3zero}
\delta\chi
&=
F_{MN}\Gamma^{MN}\,\varepsilon 
+
\mathcal{O}(\alpha'^2) \;.
\end{align}
Here, $\Gamma^M$ satisfy the Clifford algebra in ten dimensions, $\mathcal{H}_M=H_{MNP}\Gamma^{N}\Gamma^{P}$, $\mathcal{H}=H_{MNP}\Gamma^{M}\Gamma^{N}\Gamma^{P}$, and $\varepsilon$ is a ten-dimensional Majorana-Weyl spinor.

Hence, a supersymmetric solution of the theory, neglecting terms of order ${\alpha'}^2$ and higher, satisfies
\begin{align}
\label{eq:SUSY1}
\left(
\nabla_M+\frac{1}{8}\mathcal{H}_M 
\right)\,\varepsilon 
&=0
\\
\label{eq;SUSY2}
\left(\slashed\partial\phi+\frac{1}{12}\mathcal{H} 
\right)\,\varepsilon 
&=0
\\
\label{eq:SUSY3}
F_{MN}\Gamma^{MN}\,\varepsilon 
&=
0
\;.
\end{align}

Let us conclude this section with a few remarks on an integrability result and the different connections that appear in the action, Bianchi identity and supersymmetry conditions. Note first that \eqref{eq:SUSY1} can be written as
\begin{align}
\label{eq:SUSY1Bismut}
\nabla^{+}_M\,\varepsilon = 0
\end{align}
where $\nabla^{+}$ is the covariant derivative of the connection 
\begin{align}\label{eq:bismutconnectiondef}
\omega_{IJ}^{+\;\;K}=\omega_{IJ}^{\;\;\;\;K}+\frac{1}{2}H_{IJ}^{\;\;\;\:K}\;,
\end{align}
where $\omega$ is again the Levi-Civita connection. The connection $\omega^+$ is commonly referred to as \emph{Bismut connection} in the literature. 

Hence, we encounter two different connections in action and Bianchi identity on the one hand and supersymmetry conditions on the other hand. This leads to an integrability result which was first derived in Ref.~\cite{Ivanov:2009rh}. An alternative derivation using spinor methods can be found in Ref.~\cite{Martelli:2010jx}. The integrability result states that the supersymmetry conditions imply the equations of motion if and only if the connection $\omega^-$ satisfies
\begin{align}\label{integrabilitycondition}
R^-_{MNKL}\Gamma^{MN}\varepsilon = 0\;.
\end{align}
It can be shown~\cite{Martelli:2010jx} in general that this condition is automatically satisfied up to corrections of first order in $\alpha'$. This means that a field configuration which solves the supersymmetry conditions \eqref{eq:SUSY1zero}-\eqref{eq:SUSY3zero} and the Bianchi identity \eqref{eq:bianchi} ignoring all $\mathcal{O}(\alpha')$ terms solves the equations of motion derived from the action \eqref{eq:action}, again ignoring all terms $\mathcal{O}(\alpha')$. To see this, denote by $
R^{\pm\,(0)}$ and $H^{(0)}$ solutions to the supersymmetry conditions and Bianchi identity ignoring $\mathcal{O}(\alpha')$ corrections, so that, in particular, $dH^{(0)} = 0$. However, from the definition of the connections $\omega^\pm$ we have (note the index structure)
\begin{align}
R^{+\,(0)}_{KLMN} - R^{-\,(0)}_{MNKL}=\frac{1}{2}(dH^{(0)})_{KLMN}
\end{align}
and, therefore, equality of the two curvature forms, $R^{+\,(0)}_{KLMN} = R^{-\,(0)}_{MNKL}$, follows. Combing this with 
\begin{align}
[\nabla^+_M , \nabla^+_N]\,\varepsilon = R^+_{MNKL}\Gamma^{KL}\,\varepsilon = 0\; ,
\end{align}
a direct conclusion from the gravitino variation~\eqref{eq:SUSY1Bismut}, the integrability condition \eqref{integrabilitycondition} follows. This argument may, of course, break down at order $\alpha'$ since the flux need not be closed. For our purposes, it is sufficient that the integrability condition is satisfied to lowest order. This guarantees that the equations of motion are satisfied to order $\alpha'$, the order we are working to in this paper, provided, of course, Killing spinor equations and Bianchi identity are satisfied to the same order \cite{Martelli:2010jx}.

\subsection{Heterotic half--BPS domain wall solutions}
Our 10-dimensional solutions consist of a six-dimensional space with $SU(3)$ structure (the half-flat mirror or, more specifically, nearly K\"ahler spaces) and a four-dimensional domain wall, as described in Refs.~\cite{Lukas:2010mf, Klaput:2011mz}. This amounts to choosing the $1 + 2$ dimensions along the domain wall to be maximally symmetric and the remaining seven dimensions to form a non-compact $\Gtwo$-structure manifold. The associated metric takes the form
\begin{align}\label{domainwallansatz}
\d s^2=\eta_{\alpha\beta}\d x^\alpha\d x ^\beta+\underbrace{\d y^2+\underbrace{g_{uv}(x^m)\d x^u\d x^v}_{\text{$X,\;\mathrm{SU}(3)$ structure}}}_{\text{$Y,\;\mathrm{G}_2$ structure}}\;.
\end{align}
Here $\alpha$, $\beta$, ... range from 0 to 2 and label the domain wall coordinates, $y=x^3$ is the remaining four-dimensional direction, transverse to the domain wall, and $u$, $v$, ... run from 4 to 9 and label coordinates of the internal compact manifold $X$. The indices $m$, $n$, ... run from 3 to 9 and label all seven directions of the $G_2$ structure manifold $Y$. 

As evident from the above equation, the seven dimensional $\mathrm{G}_2$ structure manifold $Y$ is a warped product of the $y$ direction and the $\mathrm{SU}(3)$ structure manifold $X$. To describe this structure mathematically, it is most convenient to formulate the $\mathrm{G}_2$ and $\mathrm{SU}(3)$ structures in terms of differential forms, which we will do in the next section.

\subsection{$\mathrm{G}_2$ and $\mathrm{SU}(3)$ structure from the supersymmetry conditions}

We now briefly review how the conditions for unbroken supersymmetry,  \eqref{eq:SUSY1zero}-\eqref{eq:SUSY3zero}, give rise to the $\mathrm{G}_2$ and $\mathrm{SU}(3)$ structures of the domain wall solution \eqref{domainwallansatz}, mainly following Ref.~\cite{Lukas:2010mf}.

The general ten dimensional Majorana-Weyl spinor $\varepsilon$ which appears in the supersymmetry conditions \eqref{eq:SUSY1zero}-\eqref{eq:SUSY3zero} is decomposed in accordance with our metric Ansatz \eqref{domainwallansatz} as
\begin{align}
\varepsilon(x^m)=\rho\otimes\eta(x^m)\otimes\theta\;.
\end{align}
Here $\theta$ is an eigenvector of the third Pauli matrix $\sigma^3$, $\eta(x^m)$ is a seven dimensional spinor, and $\rho$ is a constant Majorana spinor in 2+1 dimensions and represents the two preserved supercharges of the solution. Hence, from the viewpoint of four-dimensional $N=1$ supergravity, the solution is $\frac{1}{2}$--BPS.

The spinor $\eta(x^m)$ can be used to define a three-form
\begin{equation}
\varphi_{mnp}=-i\eta^\dagger\gamma_{mnp}\eta
\end{equation}
and a four-form
\begin{equation}
\Phi_{mnpq}=\eta^\dagger\gamma_{mnpq}\eta
\end{equation}
where $\gamma_{m\dots n}:=\gamma^m\dots\gamma^n$ is a product of seven dimensional Dirac matrices. The two forms $\varphi$ and $\Phi$ define a $\Gtwo$-structure and are both Hodge dual to each other with respect to the metric $g_7=\d y^2+g_{uv}(x^m)\d x^u\d x^v$, that is, $\varphi=*_7 \Phi$. Therefore, this is the metric compatible with the so defined $\Gtwo$-structure on $\{y\} \ltimes X$ \cite{Joyce}.

Now, it can be shown that the first two supersymmetry conditions\footnote{Together with the requirement that the $H$-flux has only legs in the compact directions.} \eqref{eq:SUSY1} and \eqref{eq;SUSY2} are satisfied if and only if \cite{Lukas:2010mf, Gran:2005wf, Gauntlett:2002sc, 2003JGP....48....1F}
\begin{align}
\label{eq:killing1}
\d_7\varphi&=2\d_7\phi\wedge\varphi-*_7H 
\\
\d_7 *_7\varphi&=2\d_7\phi\wedge*_7\varphi
\\
\varphi\wedge H&=2*_7\d_7\phi\;,\\
\label{eq:killing4}
*_7\varphi\wedge H&=0 \;.
\end{align}
Here, $*_7$ is the seven-dimensional Hodge-star with respect to the metric $g_7$ and and $\d_7=dx^m\partial_m$ is the seven-dimensional  exterior derivative.
\\\\
To focus on the compact space $X$, we will now decompose these equations by performing a $6+1$ split. The forms $\varphi$ and $\Phi$ can be written in terms of six dimensional forms as
\begin{align}\label{phi:Hitchinflow}
\varphi&=-\d y\wedge J+\Omega_-\\
*_7\varphi&=\d y\wedge\Omega_++\frac{1}{2}J\wedge J\;,
\end{align}
where $J$ is a two-form and $\Omega=\Omega_++\I\,\Omega_-$ a complex three-form which, together, define an $\SU{3}$-structure on $X$.
In terms of these forms, Eqs.~(\ref{eq:killing1})-(\ref{eq:killing4}), can be re-written as
\begin{align}\label{eq:killingspinor:1}
\d\Omega_-&=2\d\phi\wedge\Omega_-
\\\label{eq:killingspinor:2}
\d J&=2\partial_y\phi\,\Omega_--\partial_y\Omega_--2\d\phi\wedge J+*H
\\\label{eq:killingspinor:3}
J\wedge\d J&=J\wedge J\wedge\d\phi
\\\label{eq:killingspinor:4}
\d\Omega_+&=J\wedge\partial_yJ-\partial_y\phi\, J\wedge J+2\d\phi\wedge\Omega_+
\\\label{eq:killingspinor:5}
J\wedge H&=*\d\phi
\\\label{eq:killingspinor:6}
\Omega_-\wedge H&=(2\partial_y\phi)\;*1
\\\label{eq:killingspinor:7}
\Omega_+\wedge H&=0
\end{align}
where all symbols and forms are quantities on the six-dimensional compact internal space $X$. In particular, $*$ denotes the six-dimensional Hodge dual with respect to the metric $g_6=g_{uv}(x^m)\d x^u\d x^v$.

An $\SU{3}$ structure can be characterised by the decomposition of the torsion tensor into irreducible $\SU{3}$ representations, as reviewed in Appendix \ref{app:su3}. The structure decomposes into five torsion classes, which are related to the exterior derivatives of $J$ and $\Omega$ via \eqref{eq:dJ} and \eqref{eq:dO}. Using these relations, it can be shown that the supersymmetry conditions \eqref{eq:killingspinor:1}-\eqref{eq:killingspinor:7} restrict the torsion classes to
\begin{align}\label{torsionclasses:genhf}
W_1^-=0 \qquad  W_2^-=0  \qquad W_4=\d\phi \qquad W_5=2\d\phi\;
\end{align}
and the remaining classes arbitrary. For the special case $H=0$, $d\phi=0$ this means that all but $W_1^+$ and $W_2^+$ vanish and such \SU{3} structures are referred to as \emph{half-flat}. Such half-flat $\SU{3}$ structures $(J,\Omega)$ can also be characterized by the relations $d\Omega_-=0$ and $J\wedge \d J=0$. Without such a restriction, \SU{3} structures satisfying \eqref{torsionclasses:genhf} are often referred to as \emph{generalised half-flat}.

Recall that the Strominger system is characterized by the stronger conditions
\begin{align}\label{torsionclasses:strominger}
W_1=0 \qquad  W_2=0  \qquad W_4=\d\phi \qquad W_5=2\d\phi\;.
\end{align}
Therefore, the Strominger system -- which results from a metric Ansatz with a maximally symmetric four-dimensional space-time -- is seen to be a special case of the more general Ansatz \eqref{domainwallansatz}, as one would have expected. Specializing \eqref{torsionclasses:strominger} further and setting $H=0$, $d\phi=0$ forces all torsion classes to vanish which corresponds to the case of Calabi-Yau manifolds times four-dimensional Minkowski space.\\[0.3cm]
In addition to the above conditions which restrict the gravitational sector of the supergravity, the instanton condition \eqref{eq:SUSY3} for a gauge field lying purely in the compact space $X$ is equivalent to the conditions
\begin{align}
\label{eq:inst1}
\Omega\neg F&=0\\
\label{eq:inst2}
J\neg F&=0\;,
\end{align}
known as the \emph{Hermitean Yang-Mills equations} (HYM). Solving these equations turns out to be a technical challenge in any heterotic compactification. For compactifications on Calabi-Yau manifolds, these are usually solved using the Donaldson-Uhlenbeck-Yau theorem which, roughly, states that every holomorphic poly-stable bundle on a compact K\"ahler manifold admits a unique Hermitean-Yang Mills connection. The geometries \eqref{torsionclasses:genhf} are in general not K\"ahler (and not even complex, since $W_1\not =0$ and $W_2\not = 0$) and, therefore, this aforementioned theorem does not apply. However, explicit solutions to the HYM equations for Abelian gauge fields on homogeneous half-flat manifolds have been obtained in Ref.~\cite{Klaput:2011mz}. Taking into account the order $\alpha'$ backreaction of these gauge fields via the Bianchi identity is one of the main purposes of this paper.

\subsection{Half-flat mirror geometry}\label{sec:mirrorgeom}

Before we move to explicit domain wall solutions on homogeneous spaces, we would like to review a convenient language in which to formulate the fundamental equations discussed in the previous section. As we have seen, the supersymmetry conditions can be cast in terms of the $\SU{3}$ structure $(J,\Omega)$, see Eqs.~\eqref{eq:killingspinor:1}-\eqref{eq:killingspinor:7}. For half-flat mirror manifolds this can be made more concrete by introducing a language analogous to Calabi-Yau manifolds. It turns out that this language also applies to the explicit examples of nearly K\"ahler coset spaces considered here~\cite{Klaput:2011mz}.

Half-flat mirror manifolds were introduced in Refs.~\cite{Gurrieri:2002wz, Gurrieri:2004dt, Gurrieri:2007jg} in the context of type II mirror symmetry with NS fluxes.  These manifolds are equipped with a set, $\{\omega_i\}$, of two-forms, and a dual set, $\{\tilde\omega^i\}$, of four forms. They also have a symplectic set, $\{\alpha_A,\beta^B\}$, of three-forms, as in the Calabi-Yau case. These forms satisfy the following relations
\begin{eqnarray}\label{eq:mirrorgeombasisrelations}
\int_X\omega_i\wedge\tilde\omega^j=\delta_i^j,\;\;\;\;\int_X\alpha_A\wedge\alpha_B=0,\;\;\;\;\int_X\beta^A\wedge\beta^B=0,\;\;\;\;\int_X\alpha_A\wedge\beta^B=\delta_A^B,
\end{eqnarray}
similar to the harmonic basis forms on a Calabi-Yau manifold. Furthermore, we define intersection numbers $d_{ijk}$ analogous to the Calabi-Yau case by writing (in cohomology)
\begin{equation}\label{eq:intersectionnumbers}
\omega_i\wedge\omega_j\equiv d_{ijk}\,\tilde\omega^k\,.
\end{equation}
In contrast to Calabi-Yau manifolds, however, these forms are not harmonic anymore in general. Instead, they satisfy the differential relations
\begin{equation}\label{eq:mirrorgeombasisrelations}
\d\omega_i =e_i\beta^0\;,\quad \d\alpha_0=e_i\tilde\omega^i\;,\quad\d\tilde\omega^i =0\;,\quad \d\beta^0=0\;.
\end{equation}
The coefficients $e_i$ are constants on $X$ and parametrize the intrinsic torsion of the manifolds. The $\SU{3}$ structure forms $J$ and $\Omega$ can be expanded in this basis
\begin{equation}
J=v^i\omega_i\; ,\quad\Omega =Z^A\alpha_A+\I\, \,G_A\,\beta^A\; ,
\end{equation}
where the fields $v^i$ are analogous to the K\"ahler moduli, the $Z^A$ analogous to the complex structure moduli and $G_A$ analogous to the derivatives of the pre-potential. Taking the exterior derivative we get
\begin{equation}\label{eq:mirrorderivative}
\d J=v^ie_i\beta^0\; ,\quad\d\Omega =Z^0e_i\tilde\omega^i\;.
\end{equation}
By comparing with Eqs.~\eqref{eq:dJ} and \eqref{eq:dO}, these results can be used to read off the torsion classes of half-flat mirror manifolds. In particular, we see that the constants $e_i$ indeed measure the intrinsic torsion of the manifolds.

The explicit construction of the above forms for the case of nearly K\"ahler coset spaces will be reviewed in the following Section and the technical details are provided in Appendix~\ref{app:cosets}.

\section{Solutions on homogeneous spaces to lowest order in $\alpha'$}\label{chap:three}

In this section we will review heterotic string solutions on coset spaces to lowest order in $\alpha'$, following Ref.~\cite{Klaput:2011mz}. This will be preparing the ground for computing the order $\alpha'$ corrections to these backgrounds in the next section.

Of the known nearly K\"ahler homogeneous spaces $\SU{3}/\U{1}^2$, $\Sp{2}/\SU{2}\times \U{1}$, $\Gtwo/\SU{3}$ and $\SU{2}\times\SU{2}$, only the first two spaces allow for line bundles using the construction method we employ. A study of the expected number of generations, using the index of the Dirac operator, shows that only $\SU{3}/\U{1}^2$ admits bundles with three generations. Hence, in our analysis we will focus on the cases $\SU{3}/\U{1}^2$ and $\Sp{2}/\SU{2}\times \U{1}$, even though the results can be extended in a straightforward way to include all four spaces. 

We will start with a brief review of coset geometry, the construction of $SU(3)$ structures on cosets and the relation to half-flat mirror geometry. Then, we discuss the construction of vector bundles and, in particular, line bundles on coset spaces. By combining these ingredients with a four-dimensional domain wall, we construct, to lowest order in $\alpha'$, 10-dimensional solutions with two supercharges to the heterotic string.

\subsection{$SU(3)$ structure on coset spaces}\label{section:cosets}
We begin with a review of coset space differential geometry and, in particular, the construction of $SU(3)$ structures. We refer to Appendix \ref{app:cosets} and Refs.~\cite{Klaput:2011mz, Castellani:1999fz} for further technical details. 

A coset space $G/H$ is obtained by identifying all elements of the Lie group manifold $G$ which are related by the action of the subgroup $H \subset G$. For the construction of bundles on $G/H$ later on, it will be useful to view $G$ as a principal fibre bundle over $G/H$ with fibre $H$, that is, $G=G(G/H,H)$. The base space $G/H$ admits a natural frame of vielbeins, which descend from the left-invariant Maurer-Cartan forms on $G$ and will be denoted by $e^i$ \cite{Castellani:1999fz}. These one-forms are, in general, no longer left-invariant under the action of $G$. However, in the cases of interest, there exist $G$-(left)-invariant two-, three- and four-forms. 

The space of G-invariant two- and three-forms for $\SU{3}/\U{1}^2$ is spanned  by~\footnote{The $G$-invariant four-forms which can be obtained from the above $G$-invariant two-forms via Hodge duality can be found in Appendix \ref{app:cosets}.}
\begin{align}\label{eq:invforms:su3def}
\{\, e^{12}\, ,\; e^{34}\, ,\; e^{56}\,\}\;,\quad \{\, e^{136}-e^{145}+e^{235}+e^{246}\,,\;  e^{135}+e^{146}-e^{236}+e^{245}\,\}\; ,
\end{align}
for $\Sp{2}/\SU{2}\times \U{1}$ by
\begin{align}\label{eq:invforms:sp2def}
\{\,e^{12}+e^{56}\,,\;e^{34}\,,\;\}\;,\quad \{e^{136}-e^{145}+e^{235}+e^{246}\,,\; e^{135}+e^{146}-e^{236}+e^{245}\,\}\; ,
\end{align}
and for $\Gtwo/\SU{3}$ by
\begin{align}\label{eq:invforms:g2def}
\{\,-e^{12}+e^{56}+e^{34}\}\;,\quad\{\,e^{136}+e^{145}-e^{235}+e^{246}\,,\; e^{135}-e^{146}+e^{236}+e^{245}\,\}\; ,
\end{align}
where $e^{i_1\dots i_n}:=e^{i_1} \wedge \dots \wedge e^{i_n}$.

Requiring the $SU(3)$ structure to be compatible with the given group structure of the coset implies that the structure forms $J$ and $\Omega$ can be expressed in terms of the above forms. Indeed, one finds that the most general $G$-invariant structure forms for $\SU{3}/\U{1}^2$ are given by
\begin{equation}\label{eq:zerothsolutionsu3}
\begin{array}{lll}
J&=&R_1^2e^{12}-R_2^2e^{34}+R_3^2e^{56}\\
\Omega&=&R_1R_2R_3\Big[\left(e^{136}-e^{145}+e^{235}+e^{246})+\I\, (e^{135}+e^{146}-e^{236}+e^{245}\right)\Big]\; ,
\end{array}
\end{equation}
with independent parameters $R_1$, $R_2$ and $R_3$. By comparing the spaces \eqref{eq:invforms:su3def} and \eqref{eq:invforms:sp2def} of $G$-invariant forms, we conclude that the most general $G$-invariant structure forms on $\Sp{2}/\SU{2}\times \U{1}$ are still given by \eqref{eq:zerothsolutionsu3} provided we set $R_1=R_3$. Similarly, the most general $G$-invariant structure on $\Gtwo/\SU{3}$ corresponds
to setting $R\equiv R_1=R_2=R_3$ in Eq.~\eqref{eq:zerothsolutionsu3}, but, in addition, with the signs of $e^2$ and $e^4$ reversed~\footnote{The sign reversal of $e^2$ and $e^4$ can be avoided by redefining the structure constants appropriately.}. This leads to
\begin{equation}\label{eq:zerothsolutiong2}
\begin{array}{lll}
J&=&-R^2e^{12}+R^2e^{34}+R^2e^{56}\\
\Omega&=&R^3\Big[\left(e^{136}+e^{145}-e^{235}+e^{246})+\I\, (e^{135}-e^{146}+e^{236}+e^{245}\right)\Big]
\end{array}
\end{equation}
for the $\SU{3}$ structures on $\Gtwo/\SU{3}$.

From the above $\SU{3}$ structure forms we can construct a unique compatible metric \cite{2001math......7101H}, which coincides with the most general $G$-invariant metric on $G/H$. For all three cases it is given by
\begin{align}\label{def:ginvmetric}
\d s^2
&=
R_1^2\,(e^1\otimes e^1+e^2\otimes e^2)
+
R_2^2\,(e^3\otimes e^3+e^4\otimes e^4)
+
R_3^2\,(e^5\otimes e^5+e^6\otimes e^6)\;
\end{align}
where for $\SU{3}/\U{1}^2$ the parameters $R_1$, $R_2$ and $R_3$ are independent, for $\Sp{2}/\SU{2}\times \U{1}$ they are restricted by $R_1=R_3$ and for $\Gtwo/\SU{3}$ by $R_1=R_2=R_3$. Hence, we recognise the parameters $R_i$ as \quot{radii} of the coset, determining the volume and shape of the space.

Having introduced $G$-invariant geometry and \SU{3} structure on our cosets, all required tools to solve the geometric sector of the heterotic string, that is, the Killing spinor equations \eqref{eq:killingspinor:1}-\eqref{eq:killingspinor:7}, are available. This has been known for some time and was first realised in Ref.~\cite{Lust:1986ix}. The additional technical difficulty of \emph{heterotic} string compactifications is the construction of vector bundles which satisfy the Hermitean Yang-Mills equations \eqref{eq:inst1}, \eqref{eq:inst2}. In past works, this has usually been approached using an Ansatz similar to the standard embedding. We will adopt the bundle construction developed in Ref.~\cite{Klaput:2011mz} which contains the standard embedding Ansatz as special case. 

\subsection{Half-flat mirror geometry of the cosets}\label{sec:mirrorgeom2}

We would now like to review the half-flat mirror geometry, in the sense of Section~\ref{sec:mirrorgeom}, for the three cosets introduced in the previous subsection. Technical details can be found in Appendix~\ref{app:cosets}. We recall that half-flat mirror geometry, in analogy with Calabi-Yau manifolds, is defined by a set of two-forms, $\{\omega_i\}$, a set of dual four-forms, $\{\tilde{\omega}^i\}$, and a set $\{\alpha_A,\beta^B\}$ of symplectic three-forms. Unlike in the Calabi-Yau case, these forms are, in general, no longer closed but instead satisfy a set of differential relations~\eqref{eq:mirrorgeombasisrelations} which involve the torsion parameters $e_i$.

It turns out that for all three cosets under consideration, there is only a single pair, $\{\alpha_0,\beta^0\}$, of symplectic three-forms in addition to a certain number of two- and four-form pairs, $\{\omega_i,\tilde{\omega}^i\}$. A subset,$\{\omega_r\}$ of the two-forms which we label by indices $r,s,\ldots$ are, in fact, closed. For $SU(3)/U(1)^2$ these forms are explicitly given by
\begin{equation}
\begin{array}{lllllll}
  \omega_1&=&-\frac{1}{2\pi}\Big(e^{12}+\frac{1}{2}e^{34}-\frac{1}{2}e^{56}\Big)&&\tilde\omega^1&=&\frac{4\pi}{3\vol}\Big(2e^{1234}+e^{1256}-e^{3456}\Big) \\
 \omega_2&=&-\frac{1}{4\pi}\Big(e^{12}+e^{34}\Big)&&\tilde\omega^2&=&-\frac{4\pi}{\vol}\Big(e^{1234}+e^{1256}\Big) \\
\omega_3&=&\frac{1}{3\pi}\Big(e^{12}-e^{34}+e^{56}\Big)&&\tilde\omega^3&=&\frac{\pi}{\vol}\Big(e^{1234}-e^{1256}+e^{3456}\Big) \\
\alpha_0&=&\frac{\pi}{2\vol}\Big(e^{136}-e^{145}+e^{235}+e^{246}\Big)&&\beta^0&=&\frac{1}{2\pi}\Big(e^{135}+e^{146}-e^{236}+e^{245}\Big)
\end{array}
\end{equation}
In particular, there are three pairs of two- and four-forms in this case. The exterior derivatives of $\omega_3$ and $\alpha_0$ are given by $d\omega_3=\beta^0$ and $d\alpha_0=\tilde{\omega}^3$, while all other forms are closed. This means the closed two-forms are $\omega_r$, where $r=1,2$. Comparing with the general differential relations~\eqref{eq:mirrorgeombasisrelations} for half-flat mirror geometry this shows that the three torsion parameters are given by $(e_1,e_2,e_3)=(0,0,1)$.

The coset $\Sp{2}/\SU{2}\times \U{1}$ has only two pairs of two- and four-forms and the explicit expressions read
\begin{equation}
\begin{array}{lllllll}
  \omega_1&=&\frac{1}{2\pi}\Big(e^{12}+2e^{34}+e^{56}\Big)&&\tilde\omega^1&=&\frac{\pi}{3\vol}\Big(e^{1234}+2e^{1256}+e^{3456}\Big) \\
  \omega_2&=&\frac{1}{6\pi}\Big(e^{12}-e^{34}+e^{56}\Big)&&\tilde\omega^2&=&\frac{2\pi}{\vol}\Big(e^{1234}-e^{1256}+e^{3456}\Big) \\
  \alpha_0&=&\frac{\pi}{2\vol}\Big(e^{136}-e^{145}+e^{235}+e^{246}\Big)&&\beta^0&=&\frac{1}{2\pi}\Big(e^{135}+e^{146}-e^{236}+e^{245}\Big)
\end{array}
\end{equation}
All but $\omega_2$ and $\alpha^0$ are closed and the non-vanishing exterior derivatives $d\omega_2=\beta^0$, $d\alpha^0=\tilde{\omega}^2$ show that the two torsion parameters are given by $(e_1,e_2)=(0,1)$. Hence, there is only one closed two-form, $\omega_1$. 

Finally, for $\Gtwo/\SU{3}$, we have
\begin{equation}
\begin{array}{lllllll}
  \omega_1&=&\frac{5}{3\pi}\Big(-e^{12}+e^{34}+e^{56}\Big)&&\tilde\omega^1&=&\frac{\pi}{5\vol}\Big(e^{1234}+e^{1256}-e^{3456}\Big)\\
  \alpha_0&=&\frac{\sqrt{3}\pi}{40\vol}\Big(e^{136}+e^{145}-e^{235}+e^{246}\Big)&&\beta^0&=&\frac{10}{\sqrt{3}\pi}\Big(e^{135}-e^{146}+e^{236}+e^{245}\Big)\; .
\end{array}
\end{equation}
In particular, there is only one pair of two- and four-forms. The non-closed forms are $\omega_1$, $\alpha_0$ with exterior derivatives $d\omega_1=\beta^0$, $d\alpha_0=\tilde{\omega}^1$ so that the single torsion parameter is $e_1=1$. Note that there is no closed two-form in this case. 

In all the expressions above, $\vol$ is the coordinate volume, a specific number whose value for each of the cosets can be found in Appendix~\ref{app:cosets}. It can be shown that the above forms indeed satisfy all the relations for half-flat mirror geometry given in Section~\ref{sec:mirrorgeom}. In particular, the $SU(3)$ structure forms on the coset spaces given in the previous subsection can be re-written in half-flat mirror form as
\begin{equation}
J = v^i\omega_i\;,\quad \Omega = Z\,\alpha_0+\I\, G\,\beta^0\; , \label{JO}
\end{equation}
where $Z$ is the single ``complex structure" modulus and $G$ the derivative of the pre-potential. From Appendix \ref{app:cosets} we see that for the first two cosets, these two quantities are related by\footnote{On $\Gtwo/\SU{3}$ the relation differs by a factor of $400/3$.}
\begin{align}\label{eq:ZGrelation}
Z = \frac{\vol}{\pi^2} G\; .
\end{align}
It is also easy to verify from the above expressions for the forms that
\begin{equation}\label{eq:mirrorsu3}
 \omega_i\wedge\alpha_0=\omega_i\wedge\beta_0=0
\end{equation}
for all $i$, in analogy with the Calabi-Yau case. These relations are also expected from the absence of $G$-invariant 5-forms on our coset spaces. A further useful relation can be deduced from the $SU(3)$ structure compatibility relation~\eqref{eq:su3}. Inserting the expansions~\eqref{JO} for $J$ and $\Omega$ into this relation leads to
\begin{equation}\label{compzeroth}
d_{ijk} v^i v^j v^k=-\frac{3}{2} ZG=-\frac{3\,\vol}{2 \pi^2}\, G^2\; .
\end{equation}
This shows that $Z$ is determined by the ``K\"ahler moduli'' $v^i$ and, therefore, no independent ``complex structure" moduli exist in our coset models.

\subsection{Levi-Civita connection}
The Levi-Civita connection is the unique torsion-free and metric compatible connection on the tangent bundle. On our spaces, with the most general $G$-invariant metric \eqref{def:ginvmetric}, the Levi-Civita connection one-form is
\begin{align}\label{def:levicevita}
{\omega^{({\rm LC})}}_b^{\phantom{b}a}=\frac{1}{2} f_{cb}^{\phantom{cb}a}e^c+f_{ib}^{\phantom{ib}a}\varepsilon^i\;.
\end{align}
The $\varepsilon^i$ are the Maurer-Cartan left-invariant one-forms on $G$ along the directions of the generators $H_i$ of the sub-group $H$. On $G/H$ these can be written in terms of the forms $e^i$, but, as we will see, the explicit expressions are not required.

The Levi-Civita connection enters the Bianchi identity \eqref{eq:bianchi} as part of the connection one-form $\omega^-$ defined in \eqref{eq:hullconnectiondef}. As we will see below, our spaces do not allow for H-flux at lowest order in $\alpha'$ and, therefore, we can set $\omega^- = \omega^{({\rm LC})}$. For $\SU{3}/\U{1}^2$ this means that the contribution to the Bianchi identity at lowest order is given as
\begin{align}\label{eq:trRRLeviCevita}
\tr\, R^{({\rm LC})}\wedge R^{({\rm LC})} = -\frac{9}{4}\,\frac{\vol}{\pi}\;\tilde{\omega}^3
\end{align}
as can be seen from Eq.~\eqref{eq:trrr:su3} (with the flux parameter $\mathcal{C}$ set to zero in this equation). The results for the other cosets can be found in Appendix~\ref{app:bianchi}. Eq.~\eqref{eq:trRRLeviCevita} will play a role when we solve the Bianchi identity iteratively,  leading us to an Ansatz for an exact solution for non-vanishing $H$ and $R^-$ in the Bianchi identity.

\subsection{Vector bundles on coset spaces}
We now turn to the problem of finding appropriate gauge bundles on the cosets, which can satisfy the Hermitean Yang-Mills equations \eqref{eq:inst1}, \eqref{eq:inst2}. Such bundles have been explicitly constructed in \cite{Klaput:2011mz}, based on the well-known relation between vector bundles and principal fibre bundle. The principal fibre bundle in our case is $G=G(G/H,H)$ and any representation $\rho\,:\,H\to\;\mathds{C}^n$ uniquely defines a rank $n$ vector bundle which is referred to as an {\em associated} vector bundle. Moreover, any connection defined on $G$ uniquely defines a connection on every associated vector bundle. We shall require the structure of the bundle to be compatible with the group structure of $G/H$. This leads to a natural connection on $G=G(G/H,H)$, related to the reductive decomposition of the Lie algebra, given by
\begin{eqnarray}
A=\varepsilon^i H_i\;.
\end{eqnarray}
Recall that $H_i$ are the generators of the Lie algebra of $H$ and the $\varepsilon^i$ are the Maurer-Cartan left-invariant one-forms on $G$ along the directions of the generators $H_i$. As before, their explicit form in terms of the vielbein $e^i$ will not be required.

On an associated vector bundle defined by the representation $\rho$, the connection associated to $A$ is then
\begin{eqnarray}
A_\rho=\varepsilon^i\rho(H_i)\;
\end{eqnarray}
with field strength
\begin{eqnarray}\label{eq:random23523}
F=-\frac{1}{2}{f_{ab}}^i\rho(H_i)e^a\wedge e^b\;.
\end{eqnarray}
Note that the one-forms $\varepsilon^i$ have indeed dropped out. This construction holds in general for every representation $\rho$ of $H$. 

We would like to add a few remarks on the \quot{standard embedding}, a choice of gauge connection frequently made in the literature. For this choice, the bundle curvature $F$ and the Riemann curvature $R$ are set equal, which solves the Bianchi identity \eqref{eq:bianchi} for $H=0$. However, in the present context, such a choice leads to a problem. Since our spaces are not Ricci-flat, the so-chosen field strength $F$ does not satisfy the Hermitian Yang-Mills (HYM) equations, so that the solution is not supersymmetric. If we choose instead
\begin{align}\label{eq:somechoice}
\rho(H_i)_b^{\phantom{b}a}=f_{ib}^{\phantom{ib}a}
\end{align}
then the curvature \eqref{eq:random23523} satisfies the HYM equations~\footnote{This choice is known as the $H$-connection on homogeneous spaces and should not be confused with the Hull connection \eqref{eq:hullconnectiondef} which is, unfortunately, often referred to as the $H$-connection as well. To avoid confusion will we not use this terminology in the present paper.}. This choice is also commonly referred to as standard embedding, even though the geometric connection and the gauge connection are not equal. 
Note that \eqref{eq:somechoice} does not solve the Bianchi identity for $H=0$ anymore. However, since this connection only differs from the Levi-Civita connection \eqref{def:levicevita} by a torsion term, both choices yield the same cohomology class for $\tr F\wedge F$ and $\tr R \wedge R$. This means that the topological constraint arising from the Bianchi identity is satisfied, while the exact identity is only satisfied to lowest order in $\alpha'$. This has been the case for most heterotic bundle constructions in past works. In contrast, we will construct exact solutions to the the Bianchi identity and solutions to order $\alpha'$ of the supersymmetry constraints.

\subsection{Line bundle sums}
When constructing a solution to the $E_8 \times E_8$ heterotic string, the structure group of a vector bundle has to be embedded in $E_8$ and the resulting low-energy gauge group will be given by the commutant of the structure group within $E_8$. Recently, it has been noted that vector bundles which consist of sums of line bundles provide a fertile class of models which can be studied systematically~\cite{Anderson:2011ns}. Such line bundle sums have been used for the half-flat compactifications in Ref.~\cite{Klaput:2011mz} and will also be the focus of the present paper.

Let us first focus on a single line bundle, $L$, defined by a one-dimensional representation $\rho\,:\,H\to\;\mathds{C}$. For $\SU{3}/\U{1}^2$, such a representation is characterized by two integers, $p^r$, where $r=1,2$, which correspond to the charges of the two $U(1)$ symmetries. Writing
\begin{align}
\rho(H_7) = -\I\,(p^1+p^2/2)\qquad \rho(H_8)=-\I\, p^2/(2\sqrt{3})\;
\end{align}
and using Eq.~\eqref{eq:random23523} the first Chern class of such a line bundle becomes
\begin{equation}
 c_1(L)=\frac{\I}{2\pi}[F]=p^r \omega_r\;.
\end{equation}
Hence, the integers ${\bf p}=(p^r)$ label the first Chern class of the line bundles and we can adopt the notation $L={\cal O}_X({\bf p})$. 
A sum of line bundles
\begin{eqnarray}
V=\bigoplus_{a=1}^n {\cal O}_X({\bf p}_a)
\end{eqnarray}
is, therefore, characterized by the set, $\{p_a^r\}$, of integers and its total first Chern class is given by
\begin{equation}
 c_1(V)=\sum_{a=1}^np_a^r \omega_r\;.
\end{equation}
The case $\Sp{2}/\SU{2}\times \U{1}$ works analogously, with each line bundle characterized by a single integer (so that $r$ only takes the value $1$ in all equations) which corresponds to the charge of the $U(1)$ factor in $H$. For $\Gtwo/\SU{3}$ the sub-group $H$ has no one-dimensional representations (except the trivial one) and no line bundles can be obtained by this construction.
Given that there are no $G$-invariant exact two-forms on our spaces, it follows that the field strength for the connection on $V$ is given by
\begin{eqnarray}
F=[F]=-2\pi i\sum_ap^r_a\omega_r\;.
\label{eq:F}
\end{eqnarray}

To ensure that the structure group of $V$ can be embedded into $E_8$, we impose the vanishing of the first Chern class, $c_1(V)=0$. This condition restricts the integers $p_a^r$ by
\begin{equation}
 \sum_{a=1}^np_a^r=0\qquad \forall\,r\;.
\end{equation}
Then, the structure group of $V$ is $S(U(1)^n)$ which is indeed a sub-group of $E_8$ for $1<n\leq 8$. Further, for $n=3,4,5$, the commutant of $S(U(1)^n)$ within $E_8$ is given by $S(U(1)^3)\times E_6$, $S(U(1)^4)\times SO(10)$ and $S(U(1)^5)\times SU(5)$, respectively. These are the phenomenologically interesting GUT gauge groups and for the ``visible" $E_8$ we will, therefore, focus on line bundle sums of rank $3$, $4$ or $5$. 

Subsequently, we will require the vector bundle contribution to the Bianchi identity~\eqref{eq:bianchi}. Focusing on the main case of interest, we evaluate this contribution for a sum of line bundles on $\SU{3}/\U{1}^2$. Writing $(p_a,q_a)=(p_a^1,p_a^2)$ for ease of notation, we find
\begin{equation}\label{eq:trFF}
\tr\, F \wedge F =- \frac{\vol}{8\pi}\left[
\sum_a(6 p_a^2 + q_a^2+6p_a q_a)\tilde{\omega}^1+
\sum_ap_a(3 p_a +2 q_a)\tilde{\omega}^2+
\frac{4}{3}\sum_a(3p_a^2 + q_a^2 + 3p_a q_a)\tilde{\omega}^3\right]
\end{equation}
Note that we will, of course, have two different bundles, one for each $E_8$ factor, corresponding to the visible and hidden sectors of the theory. Hence, the Bianchi identity has two contributions of the form~\eqref{eq:trFF}, each controlled by its own set of integers. As we will see, the hidden bundle contribution is important as it can be adjusted to cancel the other terms in the Bianchi identity.

Another basic phenomenological requirement on the visible vector bundle is the presence of three chiral generations. The number of generations is counted by the index of the bundle which can be computed using the Atiyah-Singer index theorem. For a sum of line bundles, $V$, this has been done in Appendix~\ref{app:diracindex}, leading to
\begin{equation}
 {\rm ind}(V)=-\frac{1}{6}d_{rst}\sum_{a=1}^np_a^rp_a^sp_a^t\; , \label{indV}
\end{equation}
where $d_{ijk}$ are the intersection numbers. Specializing to $\SU{3}/\U{1}^2$ gives
\begin{equation}
{\rm ind}(V)=-\sum\limits_{a=1}^{n}\left(p_a^3+\frac{1}{2}p_aq_a(q_a+3p_a)\right)\;. \label{indVsu3}
\end{equation}

\subsection{Solutions to lowest order in $\alpha'$}\label{section:lowestorder}
We have now collected all ingredients to solve the heterotic string on our coset spaces. In this section we will review the solution at lowest order in $\alpha'$ which has been found in Ref.~\cite{Klaput:2011mz}.

As discussed in Section 2, finding a supersymmetric vacuum of the heterotic string is equivalent to finding fields which satisfy the Bianchi identity \eqref{eq:bianchi}, the Killing spinor equations \eqref{eq:killingspinor:1}-\eqref{eq:killingspinor:7}, the HYM equations \eqref{eq:inst1}, \eqref{eq:inst2} and the integrability condition \eqref{integrabilitycondition}. 

The discussion below equation \eqref{integrabilitycondition} shows that the integrability condition is satisfied to lowest order. This means solving the Killing spinor equations and the Bianchi identity implies that the equations of motion are satisfied to lowest order as well.
For clarity, we will label the lowest order solution by $(0)$, except for the bundle~\footnote{We will see later that the solution at first order requires all fields to change apart from the gauge field strength.} which we will still denote by $F$. The relevant objects are then
$H^{(0)}$, $\phi^{(0)}$, $J^{(0)}$, $\Omega^{(0)}$, $g^{(0)}$ and $F$ and we will also denote the Hodge star with respect to the metric $g^{(0)}$ as $*_0$.

\subsubsection{Bianchi identity}
Let us consider the Bianchi identity first. At lowest order in $\alpha'$ it is
\begin{eqnarray}
\d H^{(0)}=0\;.
\end{eqnarray}
Now, take a look at the first two Killing spinor equations \eqref{eq:killingspinor:1}, \eqref{eq:killingspinor:2} at this order
\begin{align}
\d(e^{-2\phi^{(0)}}\Omega_-^{(0)})&=0\\
\d(e^{-2\phi^{(0)}}J^{(0)})&=-\partial_y(e^{-2\phi^{(0)}}\Omega_-^{(0)})+*_0H^{(0)}e^{-2\phi^{(0)}}.
\end{align}
Since $H^3(X)=0$ for all the spaces we are considering, these equations show that $*_0H^{(0)}e^{-2\phi^{(0)}}$ is the sum of two exact forms and, hence, an exact form itself. Using this, we have
\begin{eqnarray}
\vert\vert H^{(0)}e^{-\phi^{(0)}}\vert\vert^2=\int_{X}H^{(0)}\wedge*_0e^{-2\phi^{(0)}}H^{(0)}=0
\end{eqnarray}
after partial integration. It follows that
\begin{align}
H^{(0)}=0\;.
\end{align}
In fact, our proof holds for all domain wall compactifications on an internal manifold with $H^3(X)=0$ and, therefore, no nontrivial $H$-flux can be present in such geometries at lowest order\footnote{This result agrees with the findings of \cite{Gray:2012md}, which performed an extensive search for flux compactifications of the heterotic string on various known non-Calabi-Yau backgrounds including our cosets.}. Note that this is very similar to findings in \cite{Kimura:2006af}, which studied no-go theorems for heterotic flux compactifications with maximally symmetric four-dimensional spacetimes.

\subsubsection{Killing spinor equations}
Having solved the integrability condition and the Bianchi identity, we now turn to solving the Killing spinor equations. To lowest order the two Killing spinor equations \eqref{eq:killingspinor:5} and \eqref{eq:killingspinor:6} read
\begin{align}
0&=*_0\,\d\phi^{(0)}
\\
0&=(2\partial_y\phi^{(0)})\;*_01 \;,
\end{align}
or equivalently $\d\phi^{(0)}=\partial_y\phi^{(0)}=0$. This means that, in addition to vanishing H-flux, the dilaton is constant. The Killing spinor equations \eqref{eq:killingspinor:1}-\eqref{eq:killingspinor:7} then reduce to the Hitchin flow equations\cite{2001math......7101H}
\begin{align}\label{eq:killing:simplified}
\d\Omega_-^{(0)}&=0
\\\label{eq:killing:simplified2}
\d J^{(0)}&=-\partial_y\Omega_-^{(0)}
\\\label{eq:killing:simplified3}
J^{(0)}\wedge\d J^{(0)}&=0
\\\label{eq:killing:simplified4}
\d\Omega_+^{(0)}&=J^{(0)}\wedge\partial_y J^{(0)}
\;.
\end{align}

As can be explicitly checked, the Hitchin flow equations \eqref{eq:killing:simplified}-\eqref{eq:killing:simplified4} are solved by the $G$-invariant $\SU{3}$ structures \eqref{eq:zerothsolutionsu3}, \eqref{eq:zerothsolutiong2}, provided the parameters $R_i$ assume a certain $y$-dependence to be examined shortly.

\subsubsection{Hermitian Yang-Mills equations}\label{section:HYM}
The gauge bundle has to satisfy the equivalent of the HYM equations, that is, the instanton conditions $J\neg F=0$ and $\Omega\neg F=0$.
The second of these condition is automatically satisfied for the holomorphic three-form \eqref{eq:zerothsolutionsu3}, \eqref{eq:zerothsolutiong2} and field strengths~\eqref{eq:random23523}. The first condition, however, leads to an additional constraint on the parameters appearing in the $\SU{3}$ structure \cite{Klaput:2011mz}. To see this, note that $J\neg F=0$ is equivalent to
\begin{eqnarray}
F\wedge J\wedge J=0\;.
\label{eq:dual-inst}
\end{eqnarray}
Inserting $J=v^i \omega_i$, with the $G$-invariant two-forms $\omega_i$ and the field strength (\ref{eq:F}) into Eq.~\eqref{eq:dual-inst} gives
\begin{align}\label{eq:HYM:slope}
d_{rjk}\, p^r_av^jv^k=0\;\;\mbox{for all }\;a\; .
\end{align}
Here $d_{rjk}$ are the intersection numbers (see Appendix \ref{app:cosets}) and we recall that indices $i,j,\ldots$ run over all two-forms while indices $r,s,\ldots$ only run over the subset of closed two-forms . The solution to Eqs.~\eqref{eq:HYM:slope}, for generic values of the integers $p_a^r$, is to set all $v^r$ to zero. For $\SU{3}/\U{1}^2$ ( $\Sp{2}/\SU{2}\times\U{1}$ ) this leaves us with one remaining non-zero modulus $v^3$ ($v^2$) corresponding to the non-harmonic two-form $\d\omega_3\neq0$ ($\d\omega_2\neq0$). Therefore, from the relations \eqref{eq:moduliRSU3}-\eqref{eq:moduliRSU3last} and \eqref{eq:moduliRSP2}-\eqref{eq:moduliRSP2last} between the K\"ahler moduli $v^i$ and the radii $R_i$ we see that the HYM are solved if
\begin{align}\label{eq:allradii} 
R_1^2&=R_2^2=R_3^2\equiv R^2&\text{for }&\SU{3}/\U{1}^2&\\
R_1^2&=R_2^2\equiv R^2&\text{for }&\Sp{2}/\SU{2}\times \U{1}\;.&
\end{align}
It then follows, using the relations \eqref{eq:dJ}, \eqref{eq:dO} between the torsion classes and the \SU{3} structure forms, that the only non-vanishing torsion class is the real part of the first class $W_1^+=1/R$ \cite{1126-6708-2009-09-077, Klaput:2011mz}. This means that the $\SU{3}$ structure of $X$ is nearly K\"ahler.

There is a subtlety in the case $\SU{3}/\U{1}^2$. If $q_a=-2 p_a$ or $q_a=0$ for all $a$, the parameters $R_i$ do not have to be all equal. 
(The analogous subtlety $p_a=0$ in the case $\Sp{2}/\SU{2}\times\U{1}$ corresponds to the trivial bundle). From now on we exclude these special cases, unless otherwise stated and we will return to this possibility when we discuss the four-dimensional effective supergravity in Chapter 5. 

\subsubsection{Hitchin flow equations}\label{section:hitchinflowzero}
So far we have not determined the $y$-dependence of the $\SU{3}$ structure forms which is governed by the Hitchin flow equations (\ref{eq:killing:simplified})-(\ref{eq:killing:simplified4}). To work this out, we insert the half-flat mirror geometry expansion (which we introduced in Sections \ref{sec:mirrorgeom} and \ref{sec:mirrorgeom2}) into these flow equations. The two equations \eqref{eq:killing:simplified} and \eqref{eq:killing:simplified3} are automatically satisfied using the compatibility constraints \eqref{eq:mirrorsu3}. The other two equations become
\begin{align}
\label{eq:hitch1zeroth}
v^ie_i\,\beta^0&=-\partial_yG\;\beta^0
\\
\label{eq:hitch2zeroth}
Z\,e_i\,\tilde\omega^i&=d_{ijk}\,v^i(\partial_yv^j)\,\tilde\omega^k\;.
\end{align}
Multiplying with $\wedge (v^l \omega_l)$ on both sides of \eqref{eq:hitch2zeroth} and integrating gives
\begin{align}
Ze_k v^k=d_{ijk} v^i (\partial_y v^j) v^k\;.
\end{align}
Now, using the compatibility relation \eqref{compzeroth} we can express this in terms of the complex structure modulus
\begin{align}\label{eq:Rydependence}
e_k v^k=-\partial_y G\;,
\end{align}
which shows that equations \eqref{eq:hitch1zeroth} and \eqref{eq:hitch2zeroth} are, in fact, equivalent. We have seen previously, that the presence of the gauge fields force all radii to be equal. The $y$-dependence should, therefore, reside in this overall modulus $R=R(y)$ and we write the $SU(3)$ structure forms as
\begin{align}
J^{(0)} = R^2\,\tilde{v}^k \omega_k\; ,\qquad \Omega^{(0)}=R^3 \left(\tilde{Z}\, \alpha_0 + \I\, \tilde{G} \,\beta^0 \right)\; ,
\end{align}
with  $(\tilde{v}^k)=(0,0,\tilde{v})$ for $\SU{3}/\U{1}^2$ and $(\tilde{v}^k)=(0,\tilde{v})$ for $\Sp{2}/\SU{2}\times \U{1}$ and a constant $\tilde{v}$. The values of $\tilde{Z}$ and $\tilde{G}$ follow from this choice via Eq.~\eqref{compzeroth}. From \eqref{eq:Rydependence}, the $y$-dependence of $R$ is determined by 
\begin{align}\label{eq:ydependzerothorder}
\partial_y R = - \frac{\tilde{v}}{3\,\tilde{G}}\;.
\end{align}
Since the right-hand side of this equation is a non-zero constant the solutions for $R$ are linear in $y$ and diverging as $y\to\pm\infty$. We will see later that the $\alpha'$ corrections can remove this divergent behaviour. 

\subsection{Side issues: Kaluza-Klein gauge group and Wilson lines}
An obvious question is whether the symmetries of our coset spaces $G/H$ lead to a Kaluza-Klein gauge group in four dimensions, in addition to the remnants of the $E_8\times E_8$ gauge group. It turns out \cite{Coquereaux:1984mj} that Kaluza-Klein gauge fields from such spaces take values in the quotient $N(H)/H$ where $N(H)$ is the normaliser of $H$ in $G$. For our cosets, this quotient is merely a discrete group. For example, for $\SU{3}/\U{1}^2$, with $H=\U{1}^2$, one finds that $N(H)/H\cong S_3$, the permutation group of three elements. Hence, a Kaluza-Klein gauge group in four dimensions does not arise.

The standard method to break GUT gauge groups in heterotic constructions is to include a Wilson line in the gauge bundle. This requires a non-trivial first fundamental group of the underlying space. However, all coset spaces studied here are simply connected and, hence, do not admit any Wilson lines. Alternatively, if the space admits a freely-acting symmetry a closely related compactification can be defined on the quotient manifold which has a non-trivial first fundamental group and, hence, allows for the inclusion of Wilson lines. However, for our cosets it has been shown \cite{kobayashi1} that only torsion-free discrete groups can have a free action on $G/H$, that is, groups which do not posses any cyclic elements. In particular, this excludes all finite groups. The mathematical literature provides an existence theorem for a freely acting infinite but finitely generated discrete freely-acting group on every coset of compact groups $G$, $H$. However, we have not been able to find such a group explicitly for one of our cosets. For this reason, Wilson line breaking of the GUT symmetry is not currently an option. Instead, flux in the standard hypercharge direction might be used. Such details of particle physics model building are not the primary concern of the present paper and will not be discussed further.

\section{Solutions on homogeneous spaces including $\alpha'$ corrections}\label{chap:four}

In the previous sections we have seen how to construct lowest order solutions to the heterotic string on homogeneous spaces, using the associated vector bundle construction on cosets. It turns out that the four-dimensional space time is a domain wall and that the radius, $R$, of the internal space varies linearly with $y$, the coordinate transverse to the domain wall.

How do we expect this to change if we include first order $\alpha'$ corrections? In our discussion before, we saw that the Bianchi identity \eqref{eq:bianchi} at lowest order requires the three-form flux $H$ to be closed, which forces $H$ to vanish at lowest order. Now, at the next order the Bianchi identity is
\begin{align}\label{eq:bianchifull}
\d H=\frac{\alpha'}{4}\left(\tr F\wedge F-\tr R^{-}\wedge R^{-}\right)
\end{align}
and we expect a non-zero $H$ which is not closed. From a four-dimensional point of view, flux will contribute to the (super)-potential and we, therefore, expect some effect on moduli. Of course, the non-zero $H$ also feeds into the gravitino and dilatino Killing spinor equations and will change the gravitational background. 

In order to work this out, we first need to find solutions to the Bianchi identity \eqref{eq:bianchifull} and then solve the Killing spinor equations \eqref{eq:killingspinor:1}-\eqref{eq:killingspinor:7}, the Hermitean Yang-Mills equations \eqref{eq:inst1}, \eqref{eq:inst2} and the integrability condition \eqref{integrabilitycondition}. Of those, only the Bianchi identity and the integrability condition are changed by $\alpha'$ effects.

\subsection{Perturbative solution}
We begin by solving the Bianchi identity \eqref{eq:bianchifull} iteratively, using the lowest order solution on the right-hand side, in order to get an intuition for what form the general solutions will take. For concreteness, let  us perform the analysis on $\SU{3}/\U{1}^2$. The results for the other cases are summarized in Appendix \ref{app:cosets}. Taking the quantities on the right-hand side of Eq.~\eqref{eq:bianchifull} to be at zero in $\alpha'$ we can write
\begin{align}\label{firstorderHbianchi}
\d H^{(1)} = \frac{\alpha'}{4}\left(\tr F\wedge F-\tr R^{({\rm LC})}\wedge R^{({\rm LC})}\right)\; .
\end{align}
The explicit expressions for the terms in the bracket have already been computed in Eqs.~\eqref{eq:trFF} and \eqref{eq:trRRLeviCevita}. However, the gauge field contribution, $\tr F\wedge F$, includes both $E_8$ sectors so we should add two terms of the form~\eqref{eq:trFF}, one for the observable sector with bundle parameters $p_a$, $q_a$, where $a=1,\ldots ,n$, and one for the hidden sector with bundle parameters $\tilde{p}_a$, $\tilde{q}_a$, where $a=1,\ldots ,\tilde{n}$.

An integrability condition for the Bianchi identity~\eqref{firstorderHbianchi} is that the right-hand side is trivial in cohomology. Noting that $\tilde \omega^3 =\d\alpha^0$, we see from \eqref{eq:trRRLeviCevita} that $\tr R^{({\rm LC})}\wedge R^{({\rm LC})}$ is already cohomologically trivial and, hence, the same should be required for $\tr F\wedge F$. This leads to relations between the observable and hidden bundle parameters which can be written as
\begin{align}\label{eq:constraintpq1}
\sum\limits_{a=1}^{n}(6p_a^2 + q_a^2+6p_a q_a)+\sum\limits_{a=1}^{\tilde{n}}(6\tilde p_a^2 + \tilde q_a^2+6\tilde p_a \tilde q_a)&=0
\\\label{eq:constraintpq2}
\sum\limits_{a=1}^{n}p_a(3p_a +2q_a)+\sum\limits_{a=1}^{\tilde{n}}\tilde{p}_a(3\tilde p_a +2\tilde q_a)&=0\;.
\end{align}
Clearly solutions to these equations exist and explicit examples will be considered later. Note that the presence of the hidden bundle is helpful in that is can be used to cancel the observable bundle contributions which may be somewhat constrained by model building considerations.

Assuming we have satisfied the above constraints, the Bianchi identity takes the form
\begin{align}\label{firstorderHbianchi2}
\d H^{(1)} = \frac{\vol}{\pi}\operatorname{\mathcal{B}}(\mathbf{p},\mathbf{q}, \tilde{\mathbf{ p}}, \mathbf{\tilde q})\,\alpha'\, \d\alpha_0
\;.
\end{align}
with some function $\operatorname{\mathcal{B}}$ of the bundle parameters whose specific form is not important for now and will be stated later.
Using that $H^3(X)=0$ for all our spaces, we can immediately integrate this equation and obtain\footnote{To be more precise, $H$ may still contain an exact (non-$G$-invariant) piece. However, from \eqref{eq:susycond3} it follows that $\d\phi=0$, which together with \eqref{eq:susycond2} and \eqref{eq:susycond4} implies that this exact piece has to be zero.}
\begin{align}\label{firstorderH}
H^{(1)} = \frac{\vol}{\pi}\operatorname{\mathcal{B}}(\mathbf{p},\mathbf{q}, \tilde{\mathbf{ p}}, \mathbf{\tilde q})\,\alpha'\, \alpha_0
\;.
\end{align}
Even though this was evaluated for $\SU{3}/\U{1}^2$ the result is similar for the other cosets, although the precise form of $\operatorname{\mathcal{B}}$ depends on the coset.

What back-reaction does this have on the geometry of the homogeneous spaces? This can be seen from the Killing spinor equations  \eqref{eq:killingspinor:1}-\eqref{eq:killingspinor:7}, which we repeat for convenience.
\begin{align}\label{eq:susycond1}
\d\Omega_-&=2\d\phi\wedge\Omega_-
\\\label{eq:susycond2}
\d J&=2\partial_y\phi\Omega_--\partial_y\Omega_--2\d\phi\wedge J+*H
\\\label{eq:susycond3}
J\wedge\d J&=J\wedge J\wedge\d\phi
\\\label{eq:susycond4}
\d\Omega_+&=J\wedge\partial_yJ-\partial_y\phi J\wedge J+2\d\phi\wedge\Omega_+
\\\label{eq:susycond5}
J\wedge H&=*\d\phi
\\\label{eq:susycond6}
\Omega_-\wedge H&=(2\partial_y\phi)\;*1
\\\label{eq:susycond7}
\Omega_+\wedge H&=0
\;.
\end{align}
From the orthogonality of the forms $\alpha_0$ and $\beta^0$, Eq.~\eqref{eq:mirrorgeombasisrelations}, we see that the solution \eqref{firstorderH} automatically solves Eq.~\eqref{eq:susycond7}. Using the orthogonality of $\omega_i$ and $\alpha_0$, Eq.~\eqref{eq:mirrorsu3}, Eq.~\eqref{eq:susycond5} immediately leads to $\d\phi=0$ and, hence, conditions \eqref{eq:susycond1}, \eqref{eq:susycond3} reduce to $\d\Omega_-=0$ and $J\wedge\d J = 0$. These are exactly the same relations as obtained at zeroth order in $\alpha'$ (see Eqs.~\eqref{eq:killing:simplified} and \eqref{eq:killing:simplified3}). This means that the internal geometry remains half-flat even after switching on $\alpha'$ corrections. However, from \eqref{eq:susycond6} we see that now $\partial_y\phi\not = 0$, which will impact on the Hitchin flow equations \eqref{eq:susycond2} and \eqref{eq:susycond4}, leading to a different $y$ dependence of $R$. If we were now to proceed to the second order in $\alpha'$, it seems likely that the right-hand side of the Bianchi identity at the next order only picks up $G$-invariant terms. Since $\alpha_0$ is the only non-closed $G$-invariant three form on all cosets, this forces $H^{(2)} \propto \alpha_0$, thereby keeping the geometry half-flat at the second and only altering the functional form of $\phi(y)$ and $R(y)$. It seems this process can be iterated, leading to an all order in $\alpha'$ solution to the Bianchi identity, which preserves the half-flat geometry of the cosets but induces higher order contributions to $\phi(y)$, $R(y)$. We will now verify that this expectation is indeed correct.

\subsection{Full solution Ansatz}
Motivated by the above discussion, we start with the following Ansatz 
\begin{equation}\label{eq:Ansatz:hflux}
\begin{array}{lll}
J&=&R(y)^2\, \tilde{v}^i\omega_i\\
\Omega&=&R(y)^3 \left(\tilde{Z}\alpha_0+\I\,\tilde{G}\,\beta^0\right)\\
H &=& \mathcal{C}(R, \alpha')\,\frac{\vol}{\pi}\,\alpha_0\\
\phi &=&\phi(y)
\end{array}
\end{equation}
for $\{(J,\Omega), H, \phi\}$. The bundle is defined to be the same as at lowest order since the only $\alpha'$ effects on $J$ and $\Omega$ are through the radius $R(y)$ which does not affect the HYM equations. The function $\mathcal{C}(R,\alpha')$ in the Ansatz for $H$ also depends on the bundle parameters and, along with $R(y)$, it has to be determined for a full solution. The tilded parameters have been defined in Section~\ref{section:hitchinflowzero}. In the following, we present explicit expressions for the space $\SU{3}/\U{1}^2$. The solutions for $\Sp{2}/\SU{2}\times \U{1}$ can be found in Appendix~\ref{app:bianchi}.

\subsection{Exact solution to the Bianchi identity}
Now, we will show that our Ansatz solves the full Bianchi identity
\begin{align}
\d H=\frac{\alpha'}{4}(\tr F\wedge F-\tr R^{-}\wedge R^{-})
\end{align}
for a particular choice of $\mathcal{C}(R, \alpha')$. For this, we need to compute $\tr R^{-}\wedge R^{-}$ where $R^{-}$ is the curvature two-form of the Hull connection
\begin{align}
\omega_{ab}^{-\;\;c}=\omega_{ab}^{\;\;\;\;c}-\frac{1}{2}H_{ab}^{\;\;\;\;c}\;.
\end{align}
On the coset $\SU{3}/\U{1}^2$ we then obtain (see Appendix \ref{app:trRR} for details and results for the other cosets)
\begin{align}
\tr R^{-}\wedge R^{-} = -\frac{3}{4}\left(3-\frac{\mathcal{C}^2}{R^4} + 2 \frac{\mathcal{C}}{R^2}\right)\frac{\vol}{\pi}\,\d\alpha_0
\;.
\end{align}
In the limit $\mathcal{C}\to 0$ we recover the zeroth order result \eqref{eq:trRRLeviCevita}, as we should. For $\tr F\wedge F$ we get the same result~\eqref{eq:trFF} as before. Including observable and hidden sector and assuming that the integrability conditions \eqref{eq:constraintpq1} and \eqref{eq:constraintpq2} are satisfied it can be written as
\begin{align}
\tr F\wedge F = \mathcal{A}(\mathbf{p},\mathbf{q}, \tilde{\mathbf{ p}}, \mathbf{\tilde q})\,\frac{\vol}{\pi}\,\d\alpha_0
\end{align}
where 
\begin{align}\label{eq:Aparameter2}
\mathcal{A}(\mathbf{p},\mathbf{q}, \tilde{\mathbf{ p}}, \mathbf{\tilde q})
=-\frac{1}{12}\left[\sum\limits_{a=1}^{n}q_a^2+\sum\limits_{a=1}^{\tilde{n}}\tilde q_a^2\right] \;.
\end{align}
It may seem that this only depends on the bundle parameters $q_a,\,\tilde q_a$, but not on $p_a,\,\tilde p_a$. However, note that this result only hold for consistent bundles satisfying the integrability conditions \eqref{eq:constraintpq1} and \eqref{eq:constraintpq2}, which relate $p_a$, $\tilde{p}_a$ with $q_a$, $\tilde q_a$.

With these results, the Bianchi identity reduces to a quadratic equation for $\mathcal{C}$ given by
\begin{align}
\mathcal{C} = \frac{\alpha'}{4}\left(\mathcal{A} +\frac{3}{4}\left(-\frac{\mathcal{C}^2}{R^4}+2\frac{\mathcal{C}}{R^2}+3\right)\right)
\;.
\end{align}
Its positive solution is\footnote{Note that there also exists a solution of the Bianchi identity for vanishing flux ($\mathcal{C}=0$), which would lead to an additional constraint on the bundle parameters. However, we are interested in solutions with flux and will not explore this solution further.}
\begin{align}
\mathcal{C}(R, \alpha') = \frac{1}{3\alpha' / 8 R^4} \left(-1 + \frac{3}{8}\frac{\alpha'}{R^2} + \sqrt{1 - \frac{3}{4} \frac{\alpha'}{R^2} + \frac{3\mathcal{A}+9}{16}\, \frac{\alpha'^2}{R^4}}\right) \label{Cexact}
\;.
\end{align}
In the large radius limit, $\frac{\alpha'}{R^{2}}\ll 1$, this function behaves as
\begin{align}\label{eq:squarerootexpansion}
\mathcal{C}(R, \alpha') = 
\left[\mathcal{B}+\frac{1}{128}\Big(27+12\mathcal{A}\Big)\,\frac{\alpha'}{R^2}-
\frac{3}{4096}\Big(-27+24\mathcal{A}+16\mathcal{A}^2\Big)\frac{\alpha'^2}{R^4}+
\mathcal{O}\left(\frac{\alpha'^3}{R^{6}}\right)\right]\,\alpha'\;.
\end{align}
In particular, we see that the flux is of order $\alpha'$ and that the proper expansion parameter is $\alpha'/R^2$, as expected. The leading term 
\begin{equation}
 \mathcal{B}=\frac{4\mathcal{A}+9}{16}\label{Bdef}
\end{equation}
is determined by the bundle parameters $\mathcal{A}$, Eq.~\eqref{eq:Aparameter2}, and is, in fact, all we will need in the following. While the above results were derived for the coset $\SU{3}/\U{1}^2$, we will express all subsequent equations in terms of $\mathcal{B}$. The case $\Sp{2}/\SU{2}\times \U{1}$ can then be obtained by setting $\mathcal{B}=1/2$, as can be seen from Appendix~\ref{app:solveBianchi}.

\subsection{Hitchin flow revisited} 

Apart from a non-vanishing $H$ and $y$-dependence of $R$, our Ansatz~\eqref{eq:Ansatz:hflux} remains unchanged from its lowest order form. This means that all equations \eqref{eq:susycond1} - \eqref{eq:susycond7} which do not contain $y$ derivatives or $H$ are automatically satisfied. 

The remaining three equations,  \eqref{eq:susycond2}, \eqref{eq:susycond4} and \eqref{eq:susycond6}, lead to differential equations for the $y$-dependence of $R(y)$ and $\phi(y)$ and inserting the Ansatz~\eqref{eq:Ansatz:hflux} into these gives
\begin{eqnarray}
R^2 \tilde v^i e_i\,\beta^0 &=&\left( 2 \partial_y\phi\, R^3\tilde G-3 R^2\partial_y R\,\tilde G\right)\beta^0+\mathcal{C} \pi \beta^0\label{eq:hitchinrev1}\\
R^3 \tilde Z e_i \tilde{\omega}^i &=&d_{ijk} \tilde{v}^i \tilde{v}^j\tilde{\omega}^k\left(2 R^3 \partial_y R -\partial_y\phi R^4\right)\label{eq:hitchinrev2}\\
-\frac{\vol}{\pi}\,\tilde{G} \mathcal{C} R^3 \alpha_0\wedge \beta^0&=&2\,\partial_y\phi\,*1\label{eq:hitchinrev3}\; .
\end{eqnarray}
A direct evaluation yields $J\wedge J\wedge J= -6\,R^6 *1$ and, therefore, Eq.~\eqref{compzeroth} yields the relation $*1 = \frac{\vol}{4\pi^2}\tilde{G}^2 \alpha_0\wedge \beta^0$. If we insert this last relation into the third flow equation \eqref{eq:hitchinrev3} and then use the result in Eq.~\eqref{eq:hitchinrev2}, we obtain 
\begin{eqnarray}
\partial_y \phi &=& -\frac{\mathcal{C}(R,\alpha')}{R^3}\label{fephi}\\
\partial_y R &=&=-\frac{1}{6 \pi}\left[\tilde v + \frac{3\pi\,\mathcal{C}(R,\alpha')}{R^2}\right]\; .\label{feR}
\end{eqnarray}
Here, we have set $\tilde{G}=2\pi$, the value appropriate for $\SU{3}/\U{1}^2$. These two equations already fully determine $R(y)$ and $\phi(y)$ and Eq.~\eqref{eq:hitchinrev1} yields no additional information. This can be seen after multiplying it with $\cdot \wedge (\tilde v^k \omega_k)$ and making use of the compatibility relation \eqref{compzeroth}, in complete analogy with the lowest order analysis in Section~\ref{section:hitchinflowzero}.

\subsection{Solving the flow equations}\label{sec:solvehitchin}
Solving the above differential equations~\eqref{fephi} and \eqref{feR} for the $y$-dependence of the radius $R$ and the dilaton $\phi$, with the function $\mathcal{C}$ from Eq.~\eqref{Cexact} inserted, leads to an exact solution of the Bianchi identity.  However, in the present paper, we are only interested in corrections up to order $\alpha'$. For this reason and to avoid unnecessary complications, we will consider these differential equations only to order $\alpha'$. Inserting the leading term in $\mathcal{C}$ from Eq.~\eqref{eq:squarerootexpansion} into Eqs.~\eqref{fephi} and \eqref{feR} leads to
\begin{align}\label{eq:alphaprimeflowexp1}
\partial_y \phi
&=
-\frac{\mathcal{B}}{R^3}\alpha'
\\\label{eq:alphaprimeflowexp2}
\partial_y R 
&= 
-\frac{1}{6 \pi}\left[\tilde v + \frac{3\pi\, \mathcal{B}\,\alpha'}{R^2}\right]
\; ,
\end{align}
where $\mathcal{B}=(4\mathcal{A}+9)/16$ for $\SU{3}/\U{1}^2$ (and $\mathcal{B}=1/2$ for $\Sp{2}/\SU{2}\times \U{1}$). The structure of the solutions to these equations depends crucially on the sign of $\mathcal{B}$ and we distinguish the three cases
\begin{equation}
\begin{array}{ll}
 \mbox{Case 1:}&\mathcal{B}=0\\
 \mbox{Case 2:}&\mathcal{B}<0\\
 \mbox{Case 3:}&\mathcal{B}>0\; .
\end{array}
\end{equation} 
Note from Eq.~\eqref{eq:Aparameter2}, that $\mathcal{B}$ is a function of the bundle parameters and that, for $\SU{3}/\U{1}^2$,  all three cases can indeed be realized for appropriate bundle choices. Let us now discuss the solution for each of these cases in turn. 

\subsubsection{Case 1, $\mathcal{B}=0$}
In this case, $H=0$, and Eqs.~\eqref{fephi} and \eqref{feR} revert to their zeroth order counterparts discussed in Section~\ref{section:hitchinflowzero}. This means that, due to a special choice of bundle, the $\alpha'$ corrections vanish and we remain with a constant dilaton and a linearly diverging radius $R$.

\subsubsection{Case 2, $\mathcal{B}<0$}\label{case2}
In this case, Eq.~\eqref{eq:alphaprimeflowexp2} allows for a special $y$-independent solution where $R$ assumes the constant value
\begin{align}\label{eq:staticR10d}
R_0^2 = \frac{3\pi\, |\mathcal{B}|\,\alpha'}{\tilde v}
\;.
\end{align}
For this static solution, the $\phi$ equation can then be easily integrated and we obtain a linear dilaton
\begin{align}\label{eq:staticR10dphi}
\phi(y) = \frac{|\mathcal{B}|}{R_0^3}\alpha'\,y
\;.
\end{align}
The behaviour of this solution is radically different from what we have seen at zeroth order. There, the radius $R$ was linearly divergent and the dilaton constant. For the above solution, this situation is reversed with $R$ constant and the dilaton linearly diverging. 

We can integrate Eq.~\eqref{eq:alphaprimeflowexp2} in general, to obtain the implicit solution
\begin{eqnarray}
y-y_0=6\pi\,\Big[-\frac{R}{\tilde v}+\frac{\sqrt{3|\mathcal{B}|\alpha'}\;}{\tilde v^{3/2}}\operatorname{arctanh}\left(\sqrt{\frac{\tilde v}{3|\mathcal{B}|\alpha'}}R\right)\Big].
\end{eqnarray}
Here $y_0$ is an arbitrary integration constant which corresponds to the position of the domain wall and will be set to zero for convenience. This solution has the generic form displayed in Fig.~\ref{fig:Rstab} (solid line) and exhibits a kink at $y=y_0=0$, indicating the position of the domain wall. It approaches the above constant solution \eqref{eq:staticR10d} for $R$ as $|y|\to\infty$, that is, far away from the domain wall. In this limit, the dilaton asymptotes the linearly divergent behaviour~\eqref{eq:staticR10dphi}.

\subsubsection{Case 3, $\mathcal{B}>0$}
\begin{figure}[h!]
\begin{center}
\includegraphics[width=90mm]{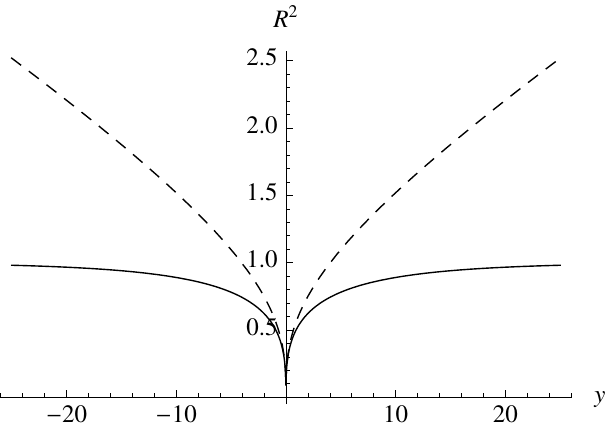}
\parbox{17.4cm}{\caption{\it Plot of the radial modulus $R^2$ as a function of the distance, $y$, from the domain wall at $y=0$ for $\mathcal{B}<0$ (solid line), and $\mathcal{B}>0$ (dashed line). For convenience, we have set $R_0^2=1$.}}
\label{fig:Rstab}
\end{center}
\end{figure}

No constant solution for $R$ exists in this case and integrating Eq.~\eqref{eq:alphaprimeflowexp2} gives
\begin{align}
y-y_0=6\pi\,\left[-\frac{R}{\tilde v}+\frac{\sqrt{3\mathcal{B}\alpha'}\;}{\tilde v^{3/2}}\operatorname{arctan}\left(\sqrt{\frac{\tilde v}{3\mathcal{B}\alpha'}}R\right)\right]
\,.
\end{align}
This solution is plotted in Fig.~\ref{fig:Rstab} (dashed line) for $y_0=0$. For $\vert y\vert \to \infty$, $R$ diverges linearly and in fact approaches the zeroth order solution \eqref{eq:ydependzerothorder}, while the dilaton becomes constant. Hence, we see that, far away from the domain wall, we recover the zeroth order solution, with a constant dilaton and a linearly divergent radius $R$.

\subsection{Discussion}
To summarize, we have seen that the qualitative behaviour of the moduli on $y$, the coordinate transverse to the domain wall, is controlled by the gauge bundle via the quantity $\mathcal{B}=(4\mathcal{A}+9)/16$ for the case of $\SU{3}/\U{1}^2$, where $\mathcal{A}$ is defined in Eq.~\eqref{eq:Aparameter2}. For $\Sp{2}/\SU{2}\times \U{1}$ there is no gauge bundle dependence and $\mathcal{B}=1/2$ always.
For $\mathcal{B}=0$ the solution is, in fact, unchanged from the zeroth order one which has a constant dilaton and a linearly divergent radius $R$. For $\mathcal{B}>0$ the solution is modified due to $\alpha'$ effects close to the domain wall but approaches the zeroth order solution far away from the domain wall. The behaviour is quite different for $\mathcal{B}<0$ which, asymptotically, leads to a constant radius $R$ and a linearly diverging dilaton.

We see that $\alpha'$ effect can have a significant effect on moduli and their stabilization. From a four-dimensional viewpoint this should be encoded in a (super-) potential which appears at order $\alpha'$. We will now discuss this in detail by considering the four-dimenional $N=1$ supergravity associated to our solutions.

\section{The four-dimensional effective theory}\label{chap:five}
Above, we have found $\mathcal{O}(\alpha')$ corrected solutions to the 10-dimensional heterotic string. In this section, we will examine the 
corresponding four-dimensional effective supergravity theories and their vacua. In particular, we would like to verify that our 10-dimensional results can be reproduced from the perspective.

\subsection{Four-dimensional supergravity and fields}
We will follow the conventions of four-dimensional supergravity laid out in Ref.~\cite{Wess:1992cp}. Mostly, we are interested in a set of chiral fields, $(\Phi^{X})$, with K\"ahler potential $K=K(\Phi^X,\bar{\Phi}^{\bar{X}})$ and superpotential $W=W(\Phi^X)$. The scalar potential is given by
\begin{eqnarray}
V=\kappa_4^{-4}e^{\kappa_4^2 K}\Big(K^{X\bar Y}F_X\bar{F}_{\bar{Y}}-3\kappa_4^2\vert W\vert^2\Big)+\frac{1}{2}\DD_a \DD^a\; ,
\end{eqnarray}
where the F-terms are defined as $F_X=\partial_XW+K_XW$, with $K_X=\partial_XK$. Further, $K_{X\bar Y}\equiv\partial_X\partial_{\bar Y}K$ is the K\"ahler metric, $K^{X\bar{Y}}$ is its inverse and $D_a$ are the D-terms. 

For compactifications on our coset spaces, the relevant moduli superfields are $(\Phi^X)=(S,T^i)$ with the dilaton $S$ and T-moduli $T^i$. We recall that the number of T-moduli depends on the specific coset. For $\SU{3}/\U{1}^2$ we have three T-moduli, so $i=1,2,3$, while $\Sp{2}/\SU{2}\times\U{1}$ has two moduli, so $i=1,2$. There are no moduli analogous to Calabi-Yau complex structure moduli.

We should now explain the relation between four- and 10-dimensional quantities, following Refs.~\cite{Gurrieri:2004dt, Gurrieri:2007jg}. First, the four-dimensional Newton constant is given in terms of its 10-dimensional counterpart by $\kappa_4^2=\kappa_{10}^2/\vol$. A set of fields, $v^i$, analogous to the K\"ahler moduli of CY manifolds, appears in the expansion
\begin{eqnarray}\label{eq:mirrorexpansion:J}
J=v^i\omega_i\;.
\end{eqnarray}
of the $\SU{3}$ structure form $J$ with respect to the two-forms $\omega_i$ of the half-flat mirror basis introduced in Sections~\ref{sec:mirrorgeom} and \ref{sec:mirrorgeom2}. We also introduce the standard quantity
\begin{equation}
 \mathcal{V}=-\frac{1}{6}d_{ijk}v^iv^jv^k\; ,
\end{equation} 
proportional to the volume of the coset space, with the intersection numbers $d_{ijk}$ explicitly given in Appendix~\ref{app:cosets}. This allows us to define the four-dimensional dilaton $s$ in terms of its 10-dimensional counterpart $\phi$ as
\begin{eqnarray}
s=e^{-2\phi}\frac{\mathcal{V}}{\vol}\; .
\end{eqnarray}
For the expansion of the 10-dimensional three-form field strength we have
\begin{eqnarray}\label{eq:hfluxmostgen}
H=-b^ie_i\beta^0+\frac{\pi^2}{\vol}\,\mu\,\alpha_0-\d b^i\wedge\omega_i+dB_4\; ,
\end{eqnarray}
where $\{\alpha_0, \beta^0\}$ is the basis of $G$-invariant three-forms introduced in Section~\ref{sec:mirrorgeom} and \ref{sec:mirrorgeom2},  $b^i$ are real scalars and $B_4$ is a two-form in four dimensions. The factor in front of the flux parameter $\mu$ is conventional in order to simplify later expressions. The first term in this expansion is due to the non-vanishing torsion of the internal space and $e_i$ are the torsion parameters. We recall that they are given by $(e_1,e_2,e_3)=(0,0,1)$ for $\SU{3}/\U{1}^2$ and $(e_1,e_2)=(0,1)$ for $\Sp{2}/\SU{2}\times\U{1}$. The second term in Eq.~\eqref{eq:hfluxmostgen} is a result of the non-vanishing H-flux induced via the Bianchi-identity. Its coefficient, $\mu$, can be read off from Eqs.~\eqref{eq:Ansatz:hflux}, \eqref{eq:squarerootexpansion} and is explicitly given by
\begin{align}\label{eq:mudefinition}
\mu=\pi\alpha'\mathcal{B}\; ,
\end{align}
where, for $\SU{3}/\U{1}^2$, the quantity $\mathcal{B}=(4\mathcal{A}+9)/16$ depends on parameters of the gauge bundle as in Eq.~\eqref{eq:Aparameter2}. For $\Sp{2}/\SU{2}\times\U{1}$ it is always given by $\mathcal{B}=1/2$. Given these preparations, we can identify the (scalar parts of the) four-dimensional superfields as
\begin{equation}
 S=a+\I  s\; ,\quad T^i=b^i+\I\, v^i\; ,
\end{equation}
where $a$ is the four-dimensional Poincar\'e-dual of the two-form $B_4$. 

\subsection{K\"ahler potential and superpotential}
The K\"ahler potential for the above set of fields is obtained from standard dimensional reduction~\cite{Gurrieri:2004dt, Gurrieri:2007jg} as
\begin{equation}
 K=-\ln\left(i(\bar{S}-S)\right)-\ln(\mathcal{V})\; .
\end{equation}
The superpotential is obtained from the generalized Gukov-Vafa-Witten formula~\cite{Gukov:1999ya,Gurrieri:2004dt}
\begin{eqnarray}\label{eq:gukovvafa}
W=-\frac{1}{Z}\int_{\tilde{X}}\Omega\wedge(H-\I\, \d J)\; .
\end{eqnarray}
After inserting the various forms from Eq.~\eqref{JO} and \eqref{eq:hfluxmostgen} and using Eq.~\eqref{compzeroth} as well as the properties of the half-flat mirror basis given in Section~\ref{sec:mirrorgeom} this leads to
\begin{equation}
 W=e_iT^i+i\mu\; .
\end{equation}
The first term arises from the non-vanishing torsion of the internal space and the second term is due to the non-vanishing H-flux induced by the gauge bundle. 

\subsection{D-terms}
The ${\rm S}(\U{1}^n)$ and ${\rm S}(\U{1}^{\tilde{n}})$ structure groups of our observable and hidden line bundle sums also appear as gauge symmetries in the four-dimensional theory. Their associated D-terms have a Fayet-Illiopoulos (FI) terms and, in general, matter field terms which involve gauge bundle moduli~\cite{Dine1987589}. Switching on these moduli deforms the gauge bundle to a one with non-Abelian structure group, a possibility which we will not consider in this paper. Focusing on the FI terms, one finds that for the observable sector
\begin{equation}\label{eq:Dterm}
 D_a\sim \frac{d_{rij}p_a^rv^iv^j}{\mathcal{V}}\; ,
\end{equation}
and similarly for the hidden sector. The D-flat conditions, $D_a=0$, hence implement the slope conditions~\eqref{eq:HYM:slope} (which follow from the HYM equations) from a four-dimensional viewpoint. Therefore, generically the D-flat conditions imply that all but the last modulus, $v=e_iv^i$, vanish as we have seen in section \ref{section:HYM}. The associated axions are absorbed by the gauge fields so we remain with a single T-modulus superfield $T=e_iT^i=b+iv$ and, of course, the dilaton $S$. In terms of these ``effective" fields the K\"ahler potential and superpotential read
\begin{equation}
 K=-\ln(S+\bar{S})-3\ln (T+\bar{T})\; ,\quad W=T+ \mu\; , \label{Weff}
\end{equation}
where we have switched to the ``phenomenological" definition $S=s+ia$ and $T=v+ib$ of the superfields, obtained from the previous one by multiplying the superfields by $-i$ and changing the signs of the axions.

It is worth noting that the above D-terms receive a dilaton-dependent correction at one loop~\cite{Lukas:1999nh,Blumenhagen:2005ga}. This correction is small in the relevant part of moduli space and will not change our conclusions, qualitatively. For simplicity, we will therefore neglect this correction.

Moreover, recall that for specific choices of the bundle parameters it is possible to satisfy \eqref{eq:Dterm} and leave more than just one of the moduli non-zero, as we pointed out at the end of section \ref{section:HYM}. However, the corresponding F-terms
\begin{align}
F_{T^s}\propto \frac{1}{\mathcal{V}}W\partial_{T^s}\mathcal{V}\propto d_{sij}v^i v^j
\;.
\end{align}
for these moduli drive the model back to the nearly-K\"ahler locus where only the last $v^i$ is non-zero. Therefore, starting from this locus covers already the most general case.

\subsection{F-term conditions}
The superpotential~\eqref{Weff} is $S$-independent and it is, therefore, expected that the dilaton cannot be stabilized. Below we will add a gaugino condensation term to $W$ in order to improve on this. However, it is still instructive at this stage to consider the F-term equations which follow from~\eqref{Weff}. For the $T$ modulus we have
\begin{equation}
 F_T=-\frac{1}{2}-\frac{3\mu}{2v}-\frac{3ib}{2v}\; .
\end{equation} 
Hence, $F_T=0$ implies a vanishing T-axion, $b=0$, and
\begin{equation}
 v=-3\mu\; . \label{v4d}
\end{equation} 
Since $v>0$ this solution is only physical provided that $\mathcal{B}<0$ and we have seen that this can be achieved for appropriate bundle choices. Indeed, this is precisely the case discussed in Section~\ref{case2} which led to a domain solution with an asymptotically constant volume given by Eq.~\eqref{eq:staticR10d}. This asymptotic value is, in fact, identical to our four-dimensional result~\eqref{v4d}, as one would expect. Of course, $F_S\sim W\neq 0$ for this value of $v$ so that we do not have a full solution to the F-term conditions but, rather, a runaway in the dilaton direction. The ``simplest" solution for this type of potential is a domain wall which is precisely what we have found previously from a 10-dimensional viewpoint.

\subsection{Including a gaugino condensate}
We will now attempt to lift the dilaton runaway by adding a gaugino condensate term to the superpotential, so that $W$ in Eq.~\eqref{Weff} is replaced by
\begin{equation}
 W=T+ \mu+ke^{-cS}\; . \label{Wnp}
\end{equation}
Here, $\mu$ is defined in Eq.~\eqref{eq:mudefinition}, $k$ is a constant of order one and $c$ is a constant depending on the condensing gauge group, with typical values  $c_{\SU{5}}=2\pi/5$, $c_{{\rm E}_6}=2\pi/12$, $c_{{\rm E}_7}=2\pi/18$ and $c_{{\rm E}_8}=2\pi/30$. In the following, it will be useful to introduce the re-scaled components
\begin{equation}
 x=cs\; ,\quad y=ca
\end{equation}
of the dilaton superfield. With those variables, the dilaton F-term equations, $F_S=0$, then read
\begin{align}
v+\mu+(1+2x)ke^{-x}\textrm{cos}(y)&=0\\
b-(1+2x)ke^{-x}\textrm{sin}(y)&=0\, ,
\end{align}
while $F_T=0$ leads to
\begin{align}
v+3\mu+3ke^{-x}\textrm{cos}(y)&=0\\
b-ke^{-x}\textrm{sin}(y)&=0\, .
\end{align}
The vanishing of the superpotential, $W=0$, is equivalent to the conditions
\begin{align}
v+\mu+ke^{-x}\textrm{cos}(y)&=0\\
b-ke^{-x}\textrm{sin}(y)&=0\, .
\end{align}
The simplest type of vacuum is a supersymmetric Minkowski vacuum, that is a solution of $F_S=F_T=W=0$. It is easy to see that this can only be achieved for $s=0$ which corresponds to the limit of infinite gauge coupling at the string scale and is, therefore, discarded. 

Next, we should consider supersymmetric AdS vacua, which are stable by the Breitenlohner-Freedman criterion. These are solutions of $F_S=F_T=0$. It follows immediately that the axions are fixed by $\cos (y)=-{\rm sign}(k)$ and $b=0$ while $x$ and $v$ are determined by
\begin{equation}
 f(x)\equiv (1-x)e^{-x}=\frac{\mu}{k}\; ,\qquad v=\frac{3x}{1-x}\mu\; .\label{min}
\end{equation} 
Normally, we require a solution with $x>1$ in order to be at sufficiently weak coupling and we will focus on this case. Then, for a positive $v$ we need the flux parameter $\mu$ to be negative and, hence, the constant $k$ to be positive. A negative value for $\mu$ is indeed possible for $\SU{3}/\U{1}^2$ but not for $\Sp{2}/\SU{2}\times\U{1}$. Provided this choice of signs, the equations~\eqref{min} have two solutions, one with a value of $x$ satisfying $1<x<2$ which is an AdS saddle and another one with $x>2$ which is an AdS minimum. The cosmological constant at those vacua is given by
\begin{equation}
 \Lambda=-\frac{3c\mu^2}{4v^3x}\left(\frac{1+x}{1-x}\right)^2\; .
\end{equation} 
We note that $v$ is stabilized perturbatively while stabilization of the dilaton involves the gaugino condensation term.
It has of course been observed some time ago~\cite{Dine:1985rz} that the dilaton in heterotic CY compactifications can be stabilized by a combination of a constant, arising from H-flux, and gaugino condensation in the superpotential. The situation here is different from these early considerations in two ways.
\begin{itemize}
 \item There is an additional T-dependent term in the superpotential which arises from the non-vanishing torsion of the internal space. 
 \item The flux term in the superpotential does not arise from harmonic H-flux but from bundle flux.
\end{itemize} 
It is important to check that the above vacuum can be in a acceptable region of field space where all consistency conditions are satisfied. To discuss this we set $\alpha'$ to one from hereon. We need that $s>1$ to be at weak coupling, $v\gg1$ so that the $\alpha'$ expansion is sensible, $k\exp (-x)<1$ so that the condensate is small and $|\Lambda|<1$ for a small vacuum energy. Eqs.~\eqref{min} immediately point to a tension in satisfying the first two of these constraints. While $v$ is proportional to the bundle flux $\mu$ and, hence, prefers a large value of $\mu$, a large value of the dilaton requires $\mu$ to be small.

Let us consider this in more detail. For concreteness we use a minimum value of $v=9$, a sufficiently large value for the $\alpha'$ expansion to be sensible. This implies the constraint
\begin{equation}
\label{eq:condmu1}
\vert\mu\vert\ge\frac{3(x-1)}{x}
\end{equation}
on the flux $\mu$. We also require the non-perturbative effects to be weak, that is $k\exp (-x)<1$, which leads to the condition
\begin{equation}
\label{eq:condmu2}
\vert\mu\vert\le x-1.
\end{equation}
Combining both conditions, it follows that $x\ge3$ and then $\Lambda<1$. Hence, the two conditions \eqref{eq:condmu1} and \eqref{eq:condmu2} are necessary and sufficient to guarantee a consistent vacuum.
\begin{figure}[h!]
\begin{center}
\includegraphics[width=85mm]{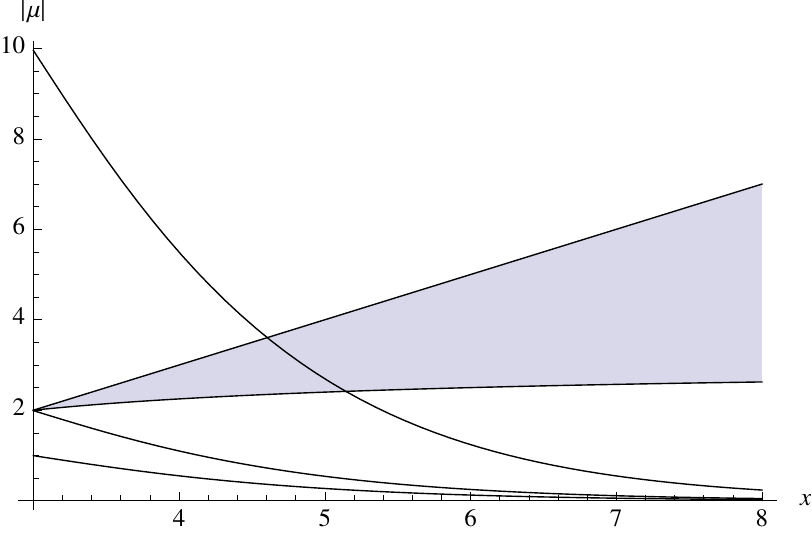}
\parbox{17.4cm}{\caption{\it Plot of the consistent values for $\vert\mu\vert$. The shaded part is defined by the conditions \eqref{eq:condmu1} and \eqref{eq:condmu2}. The other three lines represent the condition~\eqref{eq:condk} for values $k_{\rm max}=10$ (bottom line), $k_{\rm max}=20$ (middle line) and $k_{\rm max}=100$ (top line). Consistent values for the flux $|\mu|$ are, hence, defined by the shaded part located below the line for the value of $k_{\rm max}$ under consideration.}}
\label{fig:consistent}
\end{center}
\end{figure}

There is a further condition, concerning the constant $k$ in the gaugino condensation term, whose value for a given vacuum is given by 
\begin{equation}
k=\frac{\vert\mu\vert e^x}{x-1}.
\end{equation}
The general expectation is for $k$ not to be too large, so requiring it to be less than some maximum value $k_{\textrm{max}}$ implies
\begin{equation}\label{eq:condk}
\vert\mu\vert\le k_{\textrm{max}}(x-1)e^x\,.
\end{equation}
Fig.\ \ref{fig:consistent} shows the restriction on $|\mu|$ for different values of $k_{\textrm{max}}$. We see that simultaneous solutions to \eqref{eq:condmu1}, \eqref{eq:condmu2} and \eqref{eq:condk} only exist if $k_{\textrm{max}}\ge20$. For $k_{\rm max}={\cal O}(100)$ the consistent flux values are in the range $2\leq |\mu|\leq 4$. 

\subsection{Supersymmetric AdS example}
We would now like to show that the required values for the flux can indeed be obtained for appropriate choices of the gauge bundle. 
On the coset $\SU3/\U1^2$ we choose observable and hidden line bundle sums defined by the parameters
\begin{align}
(p_i)&=(-2,0,0,0,2) & (q_i)&=(1,-2,1,2,-2)\notag\\
(\tilde{p}_i)&=(2,2,0,-2,-2) & (\tilde{q}_i)&=(-3,-4,-1,4,4)\notag\;.
\end{align}
For this choice, the anomaly constraints~\eqref{eq:constraintpq1} and \eqref{eq:constraintpq2} are satisfied and the chiral asymmetry in the observable sector is three. Since both line bundle sums have rank five the gauge group in both sectors is $\mathrm{S}(\U{1}^5)\times\SU{5}$.
Computing the flux $\mu=\pi{\mathcal B}$ from Eq.~\eqref{Bdef} for this bundle choice leads to 
\begin{align}
\mu=-\frac{15\pi}{16}\approx-2.95.
\end{align}
This value is negative, as required, and indeed within the consistent range for $|\mu|$. Both the AdS saddle and the AdS minimum can be realized for this value of $\mu$, as can also be seen from Fig.~\ref{fig:susypot}. Many more consistent examples can be found on the coset $\SU3/\U1^2$. However, the situation is different for $\Sp{2}/\SU{2}\times \U{1}$. In this case, a line bundle is specified by a single integer and anomaly cancellation already fixes $\mu=\pi /2$. Since this value is positive it leads to $x<1$ so that weak coupling is difficult to achieve.

\subsection{Search for non-supersymmetric vacua}
\begin{figure}[h!]
\begin{center}
\includegraphics[width=100mm]{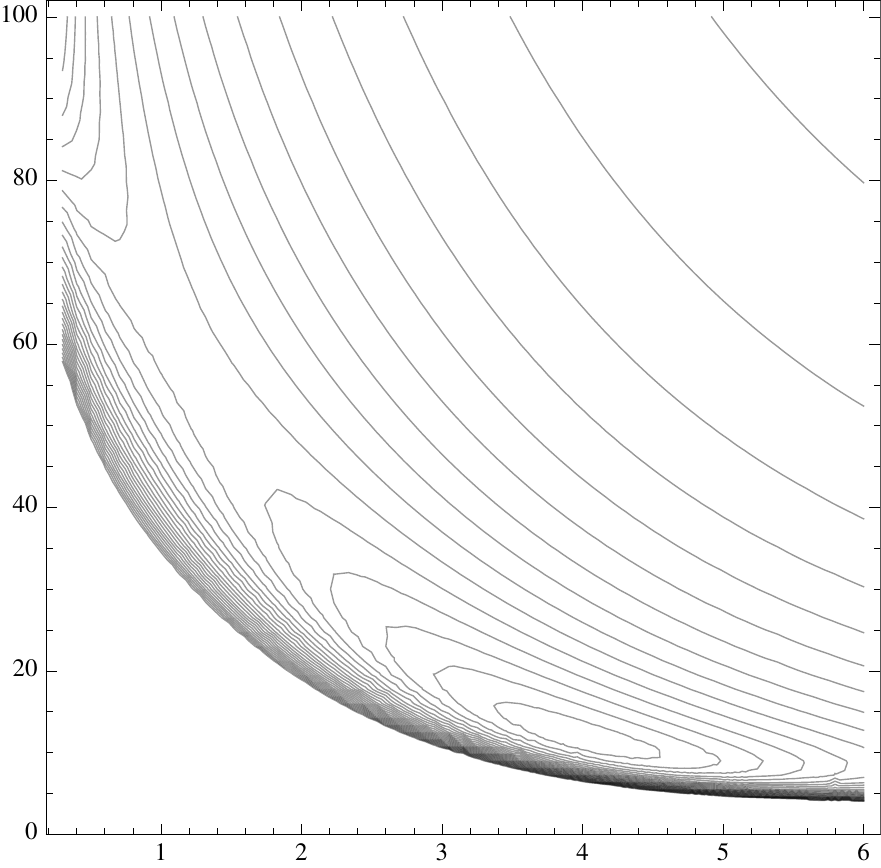}
\parbox{17.4cm}{\caption{\it Contour plot of the potential \eqref{eq:potnonpert} with $\textrm{cos}(y)=-1$ for $k=53.4$, $\mu=-15\pi /16$. This potential has a supersymmetric AdS minimum at $(x,v)\simeq (4,11.8)$, and also a supersymmetric AdS saddle at $(x,v)\simeq (1.18,58)$.}\label{fig:susypot}}
\end{center}
\end{figure}

We conclude the section by adding some remarks regarding non-supersymmetric vacua. A general search for non-supersymmetric vacua for $\SU3/\U1^2$ becomes difficult due to the presence of four complex moduli. However, one can perform an exhaustive search at the nearly K\"ahler locus -- the locus of vanishing D-terms -- where only two moduli, $S$ and $T$, remain as flat directions. On this locus, the scalar potential from the K\"ahler potential~\eqref{Weff} and the superpotential~\eqref{Wnp}, after minimizing and integrating out the axion directions, is given by
\begin{equation}
V\propto\frac{1}{sv^3}\Big(\mu^2-2v\mu-\frac{5}{3}v^2+(2x+1)^2k^2e^{-2x}\pm 2(2vx-v+\mu+2x\mu)ke^{-x}\Big)\; .
\label{eq:potnonpert}
\end{equation}
The sign of the last term equals the value of $\cos (y)=\pm 1$. A contour plot of the potential for specific values of $\mu <0$ , $k>0$ and $\cos(y)=-1$ (ensuring the existence of a supersymmetric AdS vacuum) is given in Fig.\ \ref{fig:susypot} and the two supersymmetric vacua, one AdS minimum and on AdS saddle, are clearly visible.
For the choice $k>0$ and $\cos (y)=+1$ no supersymmetric vacua exist but we find two classes of non-supersymmetric extrema, which can be both either dS and AdS, depending on the values of $k$, $\mu$. Checking the Breitenlohner-Freedman criterion, we find that all these non-supersymmetric extrema are unstable. This means that at the locus of vanishing D-terms (the nearly K\"ahler locus) only supersymmetric stable AdS vacua exist. It is still conceivable that stable non-supersymmetric vacua exist away from the nearly K\"ahler locus, but our attempts to find such vacua have remained unsuccessful. This seems to agree with recent findings in Ref.~\cite{Dolan:2012sn} where compactifications on $\SU{3}/\U{1}^2$ have been studied from a slightly different point of view.

\section{Discussion and outlook}\label{chap:six}
In this paper, we have studied heterotic domain wall compactifications on half-flat manifolds, with particular emphasis on the inclusion of $\alpha'$ corrections and moduli stabilization. In particular, we have tried to address the question as to whether the domain wall can be ``lifted" to a maximally symmetric vacuum via stabilization of all moduli. For the examples studied the answer is a cautious ``yes". A combination of $\alpha'$ and non-perturbative effects can indeed lift the runaway directions of the original, lowest-order perturbative potential and lead to a supersymmetric AdS vacuum. For appropriate bundle choices this stabilization does arise in a consistent part of moduli space, that is, at weak coupling and for moderately large internal volume. However, there is a tension in that it is not possible, for the specific examples analysed, to make the volume very large (so that there is no doubt about the validity of the $\alpha'$ expansion) and keep the theory at weak coupling. 

An explicit study of $\alpha'$ corrections and the required construction of gauge fields requires an explicit and accessible set of half-flat manifolds. For this reason, we have focused on the coset spaces which admit half-flat structures and, specifically, on $\SU{3}/\U{1}^2$ which provides the greatest flexibility among those cosets for building gauge fields via the associated bundle construction. Following Ref.~\cite{Klaput:2011mz}, we have constructed explicit gauge bundles consisting of sums of line bundles. The conditions for these gauge fields to be supersymmetric -- the D-term conditions from a four-dimensional point of view -- fix two of the three T-moduli, thereby restricting the half-flat structure to be nearly K\"ahler. We have shown that the anomaly condition can be satisfied for appropriate bundle choices and we have solved the Bianchi identity explicitly for such choices. This results in a non-harmonic H-flux, induced by the bundle flux, which leads to a correction to the metric and the dilaton profile at order $\alpha'$. These corrections preserve the nearly K\"ahler structure on the coset space. 

From a four-dimensional point of view, the bundle-induced H-flux leads to an additional, constant term in the superpotential. This term can stabilize the remaining T-modulus but the dilaton is still left a runaway direction. Upon inclusion of gaugino condensation all moduli can indeed be stabilized in a supersymmetric AdS vacuum. 

These results provide the first concrete indication that maximal symmetry at lowest order in a string solution might not be a necessary condition for a physically acceptable vacuum. This, in turn, would mean that much larger classes of internal manifolds, such as half-flat manifolds and their generalizations, are relevant in string phenomenology.  A central question in this context is, of course, how the domain wall tension, essentially set by the torsion of the manifold, can be made sufficiently small so that other effects can compete and lift the vacuum. In our examples, this can be arranged  -- at a marginal level -- by a choice of gauge bundles, although it is not possible to stabilize the theory at parametrically large volume. However, it has to be kept in mind that the coset spaces under consideration have a rather limited pattern of torsion and flux parameters available. It remains to be seen whether other half-flat manifolds offer more flexibility in this regard.\\

\noindent{\bf Acknowledgments}

\noindent We would like to thank Andrei Constantin, Sa\v{s}o Grozdanov and James Sparks for useful discussions. M.~K.\ is supported by a Lamb \& Flag scholarship of St John's College Oxford and by an STFC scholarship. A.~L.~is supported by the EC 6th Framework Programme MRTN-CT-2004-503369 and by the EPSRC network grant EP/l02784X/1. E.~E.~S.~is supported by a Oxford University Clarendon scholarship.

\appendix
\section*{\LARGE Appendix}

\section{Conventions and $\SU{3}$-structures}
In this appendix we summarize our conventions and provide a brief review of the $\SU{3}$ structure formalism and the various classes of $\SU{3}$-structure manifolds relevant to us.

\subsection{Conventions}
We decompose ten-dimensional space-time as $M_{2,1}\times\mathbb{R}\times X$, where $M_{2,1}$ is the three-dimensional Minkowski space, $\mathbb{R}$ denotes the $y$-direction transverse to the domain wall, and $X$ is the compact coset space. The index conventions are then
\begin{align}\label{indexconvention}
10d\;&:\;\;\;\;M,N,...=0,1,...,9\notag\\
7d\;&:\;\;\;\;m,n,...=3,4,...,9\notag\\
6d\;&:\;\;\;\;u,v,...=0,1,...,9\\
4d\;&:\;\;\;\;\mu,\nu,...=0,1,...,3\notag\\
3d\;&:\;\;\;\;\alpha,\beta,...=0,1,2\notag.
\end{align}
Group indices are denoted by
\begin{align}
G\;&:\;\;\;\;A,B,...=1,2,...,\textrm{dim}(G),\notag\\
G/H\;&:\;\;\;\;a,b,...=1,2,...,6\\
H\;&:\;\;\;\;i,j,...=7,8,...,\textrm{dim}(G),\notag.
\end{align}
Note that the indices $a,b,...$ correspond to the vielbein frame of the six-dimensional internal geometry labeled by the above $u,v,...$ indices.

\subsection{$\SU{3}$-structures}\label{app:su3}
A six-dimensional manifold $X$ has an $\SU{3}$-structure if there exists a real two-form $J$ and a complex three-form $\Omega$ satisfying the relations
\begin{eqnarray}
J\wedge J\wedge J=-\frac{3}{4}i\Omega\wedge\bar\Omega,\;\;\;\;\;\;\Omega\wedge J=0,
\label{eq:su3}
\end{eqnarray}
where both sides of the first equation are non-zero everywhere.

For $\d J=\d\Omega=0$, the above $\SU{3}$-structure reduces to an $\SU{3}$-holonomy for $X$. In general, $J$ and $\Omega$ are not closed and the deviation from $\SU{3}$-holonomy is measured by the intrinsic torsion $\tau$ which transforms in the $\SU{3}$ representation
\begin{eqnarray}
\tau\in(\mathbf{1}+\mathbf{1})\oplus(\mathbf{8}+\mathbf{8})\oplus(\mathbf{6}+\mathbf{\bar6})\oplus(\mathbf{3}+\mathbf{\bar3})\oplus(\mathbf{3}+\mathbf{\bar3})\; .
\end{eqnarray}
The five irreducible parts of this representation correspond to the five torsion classes $W_i$, $i=1,...,5$. They can also be explicitly read off from $\d J$ and $\d\Omega$ via the relations
\begin{align}\label{eq:dJ}
\d J&=-\frac{3}{2}\textrm{Im}(W_1\bar\Omega)+W_4\wedge J+W_3,\\\label{eq:dO}
\d\Omega&=-W_1J\wedge J+W_2\wedge J+\bar W_5\wedge\Omega,
\end{align}
where 
\begin{eqnarray}
W_3\wedge J=W_3\wedge\Omega=W_2\wedge J\wedge J=0,
\end{eqnarray}
in order for the $\SU{3}$-relations (\ref{eq:su3}) to be satisfied. For a given $\SU{3}$-structure $(J,\Omega )$ there is a unique $\SU{3}$-invariant metric $g$ and an associated almost complex structure ${\cal J}_u^v=g^{vw}J_{uw}$. This almost complex structure is integrable iff $W_1=W_2=0$. 

Some specific classes of $\SU{3}$-structures, relevant for the present paper, are characterized as follows.
\begin{align}
\textrm{nearly K\"ahler}\;\;\;\;\tau&\in W_1,\notag\\
\textrm{almost K\"ahler}\;\;\;\;\tau&\in W_2,\notag\\
\textrm{K\"ahler}\;\;\;\;\tau&\in W_5\notag,\\
\textrm{half-flat}\;\;\;\;\tau&\in W_1^+\oplus W_2^+\oplus W_3
\end{align}
where the subscript, $+$, denotes the real part of the torsion classes. Since $W_1$ and $W_2$ are non-zero the above classes of manifolds are, in general, not complex.

\section{The coset spaces}\label{app:cosets}
This appendix provides a short summary of all relevant data for the coset spaces considered in this paper, namely $\SU{3}/\U{1}^2$, $\Sp{2}/\SU{2}\times \U{1}$ and $\Gtwo/\SU{3}$. More details and derivations can be found in Ref.~\cite{Klaput:2011mz} and references therein. Although the space $\Gtwo/\SU{3}$ does not seem to allow for phenomenologically interesting models in our context, it is included for completeness. The data given here includes the generators of the Lie-group, relevant topological data and the half-flat mirror structure defined by the two-forms $\{\omega_i\}$, their four-form duals $\{\tilde\omega_i\}$ and the symplectic set $\{\alpha_0, \beta^0\}$. In accordance with our index convention \eqref{indexconvention}, the reductive decomposition of the Lie algebra of $G$ is given by $\{T_A\}=\{K_a,H_i\}$, where the $K_a$, $a=1,\dots,6$ denote the coset generators and $H_i$ the generators of the sub-group $H$.

\subsection{$\SU{3}/\U{1}^2$}
This coset is isomorphic to $\mathbb{F}^3$, the space of flags of $\mathds{C}^3$. It is also the twistor space of $\mathds{C}P^2$ and has been studied extensively in the mathematical literature. Of particular interest is the fact that it admits two almost complex structures, one of which is integrable and the other nearly K\"ahler. This is true in general for every six-dimensional manifold that is the twistor space of a four-dimensional manifold \cite{butruille}. The latter is induced by the coset structure of $\SU{3}/\U{1}^2$ and given below.

A possible choice of $\SU{3}$ generators is provided by the Gell-Mann matrices
\begin{eqnarray*}
&\lambda_1
=
-\frac{i}{2}\left(
\begin{array}{ccc}
 0 & 1 & 0 \\
 1 & 0 & 0 \\
 0 & 0 & 0
\end{array}
\right)
,\;
\lambda_2
=
\frac{1}{2}\left(
\begin{array}{ccc}
 0 & -1 & 0 \\
 1 & 0 & 0 \\
 0 & 0 & 0
\end{array}
\right)
,\;
\lambda_3
=
-\frac{i}{2}\left(
\begin{array}{ccc}
 1 & 0 & 0 \\
 0 & -1 & 0 \\
 0 & 0 & 0
\end{array}
\right)
,
\\
&\lambda_4
=
-\frac{i}{2}\left(
\begin{array}{ccc}
 0 & 0 & 1 \\
 0 & 0 & 0 \\
 1 & 0 & 0
\end{array}
\right)
,\;
\lambda_5
=\frac{1}{2}\left(
\begin{array}{ccc}
0 & 0 & -1 \\
 0 & 0 & 0 \\
 1 & 0 & 0

\end{array}
\right)
,\;
\lambda_6
=-\frac{i}{2}\left(
\begin{array}{ccc}
 0 & 0 & 0 \\
 0 & 0 & 1 \\
 0 & 1 & 0
\end{array}
\right)
,
\\
&
\lambda_7
=\frac{1}{2}\left(
\begin{array}{ccc}
  0 & 0 & 0 \\
 0 & 0 & -1 \\
 0 & 1 & 0
\end{array}
\right)
,\;
\lambda_8
=-\frac{i}{2\sqrt{3}}
\left(
\begin{array}{ccc}
 1 & 0 & 0 \\
 0 & 1 & 0 \\
 0 & 0 & -2
\end{array}
\right)
\;.
\end{eqnarray*}
The two $\U{1}$ sub-groups are generated by $\lambda_3$ and $\lambda_8$. Hence, we choose as generators the re-labelled Gell-Mann matrices
\begin{align}
\begin{aligned}
K_1&=\lambda_1\qquad K_2=\lambda_2\qquad K_3=\lambda_4\qquad K_4=\lambda_5\\
K_5&=\lambda_6\qquad K_6=\lambda_7\qquad H_7=\lambda_3\qquad H_8=\lambda_8\;.
\end{aligned}
\end{align}
The geometry of the homogeneous space $\SU{3}/\U{1}^2$ is determined by the structure constants which, relative to the basis $\{K_a, H_i\}$, are given by
\begin{eqnarray}\nonumber
&&f_{12}^{\phantom{12}7}=1\\
&&f_{13}^{\phantom{13}6}=
-f_{14}^{\phantom{14}5}=
f_{23}^{\phantom{23}5}=
f_{24}^{\phantom{24}6}=
f_{73}^{\phantom{73}4}=
-f_{75}^{\phantom{75}6}=1/2\\\nonumber
&&f_{34}^{\phantom{34}8}=
f_{56}^{\phantom{56}8}=\sqrt{3}/2
\;.
\end{eqnarray}
A basis of $G$-invariant two-, three- and four-forms is given by
\begin{equation}
\begin{array}{lllllll}
  \omega_1&=&-\frac{1}{2\pi}\Big(e^{12}+\frac{1}{2}e^{34}-\frac{1}{2}e^{56}\Big)&&\tilde\omega^1&=&\frac{4\pi}{3\vol}\Big(2e^{1234}+e^{1256}-e^{3456}\Big) \\
 \omega_2&=&-\frac{1}{4\pi}\Big(e^{12}+e^{34}\Big)&&\tilde\omega^2&=&-\frac{4\pi}{\vol}\Big(e^{1234}+e^{1256}\Big) \\
\omega_3&=&\frac{1}{3\pi}\Big(e^{12}-e^{34}+e^{56}\Big)&&\tilde\omega^3&=&\frac{\pi}{\vol}\Big(e^{1234}-e^{1256}+e^{3456}\Big) \\
\alpha_0&=&\frac{\pi}{2\vol}\Big(e^{136}-e^{145}+e^{235}+e^{246}\Big)&&\beta^0&=&\frac{1}{2\pi}\Big(e^{135}+e^{146}-e^{236}+e^{245}\Big)
\end{array}
\end{equation}
where $e^{i_1\dots i_n}:= e^{i_1} \wedge\dots\wedge e^{i_n}$ and the dimensionless volume $\vol$ is given by
\begin{eqnarray}\label{eq:VolX:su3}
\vol=\int_Xe^{123456}=4(2\pi)^3\;.
\end{eqnarray}
This G-invariant basis forms fulfil the half-flat mirror relations in Section~\ref{sec:mirrorgeom} with torsion parameters $(e_1,e_2,e_3)=(0,0,1)$ and intersection numbers
\begin{align}\label{eq:intersectionnums}
\begin{aligned}
d_{111}&=6\qquad d_{112}=3\qquad d_{113}=4\qquad d_{122}=1\qquad d_{123}=2\qquad d_{133}=0\\
d_{222}&=0\qquad d_{223}=\frac{4}{3}\qquad d_{233}=0\qquad d_{333}=-\frac{64}{9}\;.
\end{aligned}
\end{align}
The only non-zero Betti numbers are $b_0=1$, $b_2=2$, $b_4=2$ and $b_6=1$ so that the Euler number is $\chi=6$.
The most general $G$-invariant $\SU{3}$ structure forms are given by
\begin{align}
\begin{aligned}
J&=R_1^2e^{12}-R_2^2e^{34}+R_3^2e^{56}=v^i\omega_i,\\
\Omega&=R_1R_2R_3\Big((e^{136}-e^{145}+e^{235}+e^{246})+\I\, (e^{135}+e^{146}-e^{236}+e^{245})\Big)=Z\,\alpha_0+\I\, G\,\beta^0\;
\end{aligned}
\end{align}
with associated $G$-invariant metrics
\begin{align}
\d s_0^2
&=
R_1^2\,(e^1\otimes e^1+e^2\otimes e^2)
+
R_2^2\,(e^3\otimes e^3+e^4\otimes e^4)
+
R_3^2\,(e^5\otimes e^5+e^6\otimes e^6)
\;.
\end{align}
In these relations, the $R_i$ are three arbitrary ``radii" of the coset space which are related to the moduli $v^i$ by
\begin{align}\label{eq:moduliRSU3}
v^1&=-\frac{4\pi}{3}(R_1^2+R_2^2-2R_3^2)\\
v^2&=4\pi(R_2^2-R_3^2)\\
v^3&=\pi(R_1^2+R_2^2+R_3^2)\label{eq:moduliRSU3last}
\end{align}
and to $(Z,G)$ by
\begin{align}
Z=\frac{2\vol}{\pi}R_1R_2R_3\; ,\qquad
G=2\pi R_1R_2R_3\;.
\end{align}

\subsection{$\Sp{2}/\SU{2}\times \U{1}$}
As a topological space this coset is isomorphic to $\mathds{C}P^3$. Another coset realisation of $\mathds{C}P^3$ is $\SU{4}/\mathrm{S}(\U{3}\times\U{1})$ which may be more familiar to the reader. $\mathds{C}P^3$ is the twistor space of $S^4$ and, therefore, admits two almost complex structures: one integrable and the other nearly K\"ahler. The first corresponds to the invariant structure on the coset $\SU{4}/\mathrm{S}(\U{3}\times\U{1})$ while the latter corresponds to the invariant structure on $\Sp{2}/\SU{2}\times \U{1}$ and is given below.

A possible choice for the generators of the Lie-group $\Sp{2}$ is
\begin{eqnarray*}
&
K_1
=\frac{1}{\sqrt{2}}
\left(
\begin{array}{cccc}
 0 & 0 & 1 & 0 \\
 0 & 0 & 0 & 1 \\
 -1 & 0 & 0 & 0 \\
 0 & -1 & 0 & 0
\end{array}
\right),\;
K_2=\frac{i}{\sqrt{2}}\left(
\begin{array}{cccc}
 0 & 0 & 0 & 1 \\
 0 & 0 & 1 & 0 \\
 0 & 1 & 0 & 0 \\
 1 & 0 & 0 & 0
\end{array}
\right) 
,\\
&
K_3= \left(
\begin{array}{cccc}
 i & 0 & 0 & 0 \\
 0 & -i & 0 & 0 \\
 0 & 0 & 0 & 0 \\
 0 & 0 & 0 & 0
\end{array}
\right)
,
K_4
=
\left(
\begin{array}{cccc}
 0 & 1 & 0 & 0 \\
 -1 & 0 & 0 & 0 \\
 0 & 0 & 0 & 0 \\
 0 & 0 & 0 & 0
\end{array}
\right)
,
K_5
=\frac{1}{\sqrt{2}}\left(
\begin{array}{cccc}
 0 & 0 & 0 & 1 \\
 0 & 0 & -1 & 0 \\
 0 & 1 & 0 & 0 \\
 -1 & 0 & 0 & 0
\end{array}
\right)
\\
&
K_6=\frac{i}{\sqrt{2}}\left(
\begin{array}{cccc}
 0 & 0 & -1 & 0 \\
 0 & 0 & 0 & 1 \\
 -1 & 0 & 0 & 0 \\
 0 & 1 & 0 & 0
\end{array}
\right)
,\;
H_7
=
\left(
\begin{array}{cccc}
 0 & 0 & 0 & 0 \\
 0 & 0 & 0 & 0 \\
 0 & 0 & i & 0 \\
 0 & 0 & 0 & -i
\end{array}
\right)
,\\
&
H_8
=
\left(
\begin{array}{cccc}
 0 & 0 & 0 & 0 \\
 0 & 0 & 0 & 0 \\
 0 & 0 & 0 & -1 \\
 0 & 0 & 1 & 0
\end{array}
\right)
,\;
H_9=\left(
\begin{array}{cccc}
 0 & 0 & 0 & 0 \\
 0 & 0 & 0 & 0 \\
 0 & 0 & 0 & -i \\
 0 & 0 & -i & 0
\end{array}
\right)
,\;
H_{10}=\left(
\begin{array}{cccc}
 0 & i & 0 & 0 \\
 i & 0 & 0 & 0 \\
 0 & 0 & 0 & 0 \\
 0 & 0 & 0 & 0
\end{array}
\right)\;.
\end{eqnarray*}
There are two possible reductive decompositions of $\Sp{2}$ leading to two different cosets. The decomposition into $\{K_a, H_i\}$ given above corresponds to the non-maximal embedding of $\SU{2}\times \U{1}$. The other choice, the maximal embedding, leads to a different coset which does not admit a half-flat $\Sp{2}$-invariant \SU{3}-structure.

The structure constants in the given basis are
\begin{eqnarray}\nonumber
&&f_{13}^{\phantom{13}6}=
-f_{14}^{\phantom{14}5}=
f_{23}^{\phantom{23}5}=
f_{24}^{\phantom{24}6}=1\\
&&f_{71}^{\phantom{71}6}=
-f_{72}^{\phantom{72}5}=
f_{81}^{\phantom{81}5}=
f_{82}^{\phantom{82}6}=
f_{91}^{\phantom{91}2}=
-f_{95}^{\phantom{95}6}=
f_{10\;1}^{\phantom{10\;1}2}=
f_{10\;5}^{\phantom{10\;5}6}=1\\\nonumber
&&f_{78}^{\phantom{78}9}=
f_{10\;3}^{\phantom{10\;3}4}=2
\;.
\end{eqnarray}
A basis of $G$-invariant two-, three- and four-forms is given by
\begin{equation}
\begin{array}{lllllll}
  \omega_1&=&\frac{1}{2\pi}\Big(e^{12}+2e^{34}+e^{56}\Big)&&\tilde\omega^1&=&\frac{\pi}{3\vol}\Big(e^{1234}+2e^{1256}+e^{3456}\Big) \\
  \omega_2&=&\frac{1}{6\pi}\Big(e^{12}-e^{34}+e^{56}\Big)&&\tilde\omega^2&=&\frac{2\pi}{\vol}\Big(e^{1234}-e^{1256}+e^{3456}\Big) \\
  \alpha_0&=&\frac{\pi}{2\vol}\Big(e^{136}-e^{145}+e^{235}+e^{246}\Big)&&\beta^0&=&\frac{1}{2\pi}\Big(e^{135}+e^{146}-e^{236}+e^{245}\Big)
\end{array}
\end{equation}
with
\begin{equation}\label{eq:VolX:sp2}
\vol=\int_Xe^{123456}=\frac{(2\pi)^3}{12}\;.
\end{equation}
As before, these $G$-invariant forms satisfy the half-flat mirror geometry relations in Section~\ref{sec:mirrorgeom} for torsion parameters $(e_1,e_2)=(0,1)$ and intersection numbers
\begin{align}\label{eq:intersectionnumssp2}
d_{111}=1\qquad d_{112}=\frac{1}{6}\qquad d_{122}=0\qquad d_{222}&=-\frac{2}{27}\;.
\end{align}
The only non-zero Betti numbers are $b_0=b_2=b_4=b_6=1$ and, hence, the Euler number is $\chi=4$.
The most general $G$-invariant $\SU{3}$-structure forms are 
\begin{align}
\begin{aligned}
J&=R_1^2\,e^{12}-R_2^2\,e^{34}+R_1^2\,e^{56}=v^i\omega_i\\
\Omega&=R_1^2R_2\Big((e^{136}-e^{145}+e^{235}+e^{246})+\I\, (e^{135}+e^{146}-e^{236}+e^{245})\Big)=Z\,\alpha_0+\I\,G\,\beta^0\;
\end{aligned}
\end{align}
with associated $G$-invariant metrics
\begin{align}
\d s_0^2
&=
R_1^2\,(e^1\otimes e^1+e^2\otimes e^2)
+
R_2^2\,(e^3\otimes e^3+e^4\otimes e^4)
+
R_1^2\,(e^5\otimes e^5+e^6\otimes e^6)
\;.
\end{align}

The two coset radii $R_i$ are related to half-flat mirror moduli  by
\begin{align}\label{eq:moduliRSP2}
v^1&=-\frac{2\pi}{3}(R_1^2-R_2^2)\\
v^2&=2\pi(2R_1^2+R_2^2)\label{eq:moduliRSP2last}
\end{align}
and
\begin{align}
Z=\frac{2\vol}{\pi}R_1^2R_2\qquad\qquad
G=2\pi \, R_1^2R_2\;.
\end{align}

\subsection{$\Gtwo/\SU{3}$}
This coset is topologically a sphere
\begin{align}
\Gtwo/\SU{3}\cong S^6\;.
\end{align}
Like the the other spaces the sphere admits different realisations as coset, for example $\SO{7}/\SO{6} \cong S^6$. However, in contrast to the other cases there is no known integrable almost complex structure on $S^6$. The conjecture that no such almost complex structure exists is known as \emph{Chern's last theorem}. There is a well known nearly K\"ahler structure on $S^6$ which arises from the octonions (the sphere $S^6$ can be regarded as a subset of the octonions) and is invariant under the action of $\Gtwo$. This structure will be presented below.

Our choice of $\Gtwo$ generators and their reductive decomposition is
{\footnotesize
\begin{eqnarray*}
&
K_1=\frac{1}{\sqrt{3}}\left(
\begin{array}{ccccccc}
 0 & 2 & 0 & 0 & 0 & 0 & 0 \\
 -2 & 0 & 0 & 0 & 0 & 0 & 0 \\
 0 & 0 & 0 & 0 & 0 & 0 & 0 \\
 0 & 0 & 0 & 0 & 0 & 0 & 1 \\
 0 & 0 & 0 & 0 & 0 & 1 & 0 \\
 0 & 0 & 0 & 0 & -1 & 0 & 0 \\
 0 & 0 & 0 & -1 & 0 & 0 & 0
\end{array}
\right)
,\;
K_2=\frac{1}{\sqrt{3}}\left(
\begin{array}{ccccccc}
 0 & 0 & 2 & 0 & 0 & 0 & 0 \\
 0 & 0 & 0 & 0 & 0 & 0 & 0 \\
 -2 & 0 & 0 & 0 & 0 & 0 & 0 \\
 0 & 0 & 0 & 0 & 0 & 1 & 0 \\
 0 & 0 & 0 & 0 & 0 & 0 & -1 \\
 0 & 0 & 0 & -1 & 0 & 0 & 0 \\
 0 & 0 & 0 & 0 & 1 & 0 & 0
\end{array}
\right),
\\
&
K_3=\frac{1}{\sqrt{3}}\left(
\begin{array}{ccccccc}
 0 & 0 & 0 & 0 & -2 & 0 & 0 \\
 0 & 0 & 0 & 0 & 0 & 1 & 0 \\
 0 & 0 & 0 & 0 & 0 & 0 & -1 \\
 0 & 0 & 0 & 0 & 0 & 0 & 0 \\
 2 & 0 & 0 & 0 & 0 & 0 & 0 \\
 0 & -1 & 0 & 0 & 0 & 0 & 0 \\
 0 & 0 & 1 & 0 & 0 & 0 & 0
\end{array}
\right)
,\;
K_4=\frac{1}{\sqrt{3}}\left(
\begin{array}{ccccccc}
 0 & 0 & 0 & -2 & 0 & 0 & 0 \\
 0 & 0 & 0 & 0 & 0 & 0 & 1 \\
 0 & 0 & 0 & 0 & 0 & 1 & 0 \\
 2 & 0 & 0 & 0 & 0 & 0 & 0 \\
 0 & 0 & 0 & 0 & 0 & 0 & 0 \\
 0 & 0 & -1& 0 & 0 & 0 & 0 \\
 0 & -1 & 0 & 0 & 0 & 0 & 0
\end{array}
\right)
,
\end{eqnarray*}
\begin{eqnarray*}
&
K_5=\frac{1}{\sqrt{3}}\left(
\begin{array}{ccccccc}
 0 & 0 & 0 & 0 & 0 & 0 & 2 \\
 0 & 0 & 0 & 1 & 0 & 0 & 0 \\
 0 & 0 & 0 & 0 & -1 & 0 & 0 \\
 0 & -1 & 0 & 0 & 0 & 0 & 0 \\
 0 & 0 & 1 & 0 & 0 & 0 & 0 \\
 0 & 0 & 0 & 0 & 0 & 0 & 0 \\
 -2 & 0 & 0 & 0 & 0 & 0 & 0
\end{array}
\right)
,\;
K_6=\frac{1}{\sqrt{3}}\left(
\begin{array}{ccccccc}
 0 & 0 & 0 & 0 & 0 & 2 & 0 \\
 0 & 0 & 0 & 0 & 1 & 0 & 0 \\
 0 & 0 & 0 & 1 & 0 & 0 & 0 \\
 0 & 0 & -1 & 0 & 0 & 0 & 0 \\
 0 & -1 & 0 & 0 & 0 & 0 & 0 \\
 -2 & 0 & 0 & 0 & 0 & 0 & 0 \\
 0 & 0 & 0 & 0 & 0 & 0 & 0
\end{array}
\right)
,
\end{eqnarray*}
\begin{eqnarray*}
&
H_7=\left(
\begin{array}{ccccccc}
 0 & 0 & 0 & 0 & 0 & 0 & 0 \\
 0 & 0 & 0 & 0 & 0 & 0 & 0 \\
 0 & 0 & 0 & 0 & 0 & 0 & 0 \\
 0 & 0 & 0 & 0 & 0 & 0 & -1 \\
 0 & 0 & 0 & 0 & 0 & 1 & 0 \\
 0 & 0 & 0 & 0 & -1 & 0 & 0 \\
 0 & 0 & 0 & 1 & 0 & 0 & 0
\end{array}
\right)
,\;
H_8=\left(
\begin{array}{ccccccc}
 0 & 0 & 0 & 0 & 0 & 0 & 0 \\
 0 & 0 & 0 & 0 & 0 & 0 & 0 \\
 0 & 0 & 0 & 0 & 0 & 0 & 0 \\
 0 & 0 & 0 & 0 & 0 & -1 & 0 \\
 0 & 0 & 0 & 0 & 0 & 0 & -1 \\
 0 & 0 & 0 & 1 & 0 & 0 & 0 \\
 0 & 0 & 0 & 0 & 1 & 0 & 0
\end{array}
\right)
,
\end{eqnarray*}
\begin{eqnarray*}
&
H_9=\left(
\begin{array}{ccccccc}
 0 & 0 & 0 & 0 & 0 & 0 & 0 \\
 0 & 0 & 0 & 0 & 0 & 0 & 0 \\
 0 & 0 & 0 & 0 & 0 & 0 & 0 \\
 0 & 0 & 0 & 0 & -1 & 0 & 0 \\
 0 & 0 & 0 & 1 & 0 & 0 & 0 \\
 0 & 0 & 0 & 0 & 0 & 0 & 1 \\
 0 & 0 & 0 & 0 & 0 & -1 & 0
\end{array}
\right)
,\;
H_{10}=\left(
\begin{array}{ccccccc}
 0 & 0 & 0 & 0 & 0 & 0 & 0 \\
 0 & 0 & 0 & 0 & 0 & -1 & 0 \\
 0 & 0 & 0 & 0 & 0 & 0 & -1 \\
 0 & 0 & 0 & 0 & 0 & 0 & 0 \\
 0 & 0 & 0 & 0 & 0 & 0 & 0 \\
 0 & 1 & 0 & 0 & 0 & 0 & 0 \\
 0 & 0 & 1 & 0 & 0 & 0 & 0
\end{array}
\right)
,
\end{eqnarray*}
\begin{eqnarray*}
&
H_{11}=\left(
\begin{array}{ccccccc}
 0 & 0 & 0 & 0 & 0 & 0 & 0 \\
 0 & 0 & 0 & 0 & 0 & 0 & 1 \\
 0 & 0 & 0 & 0 & 0 & -1 & 0 \\
 0 & 0 & 0 & 0 & 0 & 0 & 0 \\
 0 & 0 & 0 & 0 & 0 & 0 & 0 \\
 0 & 0 & 1 & 0 & 0 & 0 & 0 \\
 0 & -1 & 0 & 0 & 0 & 0 & 0
\end{array}
\right)
,\;
H_{12}=\left(
\begin{array}{ccccccc}
 0 & 0 & 0 & 0 & 0 & 0 & 0 \\
 0 & 0 & 0 & 1 & 0 & 0 & 0 \\
 0 & 0 & 0 & 0 & 1 & 0 & 0 \\
 0 & -1 & 0 & 0 & 0 & 0 & 0 \\
 0 & 0 & -1 & 0 & 0 & 0 & 0 \\
 0 & 0 & 0 & 0 & 0 & 0 & 0 \\
 0 & 0 & 0 & 0 & 0 & 0 & 0
\end{array}
\right)
,
\end{eqnarray*}
\begin{eqnarray*}
&
H_{13}=\left(
\begin{array}{ccccccc}
 0 & 0 & 0 & 0 & 0 & 0 & 0 \\
 0 & 0 & 0 & 0 & -1 & 0 & 0 \\
 0 & 0 & 0 & 1 & 0 & 0 & 0 \\
 0 & 0 & -1 & 0 & 0 & 0 & 0 \\
 0 & 1 & 0 & 0 & 0 & 0 & 0 \\
 0 & 0 & 0 & 0 & 0 & 0 & 0 \\
 0 & 0 & 0 & 0 & 0 & 0 & 0
\end{array}
\right)
,\;
H_{14}=\frac{1}{\sqrt{3}}\left(
\begin{array}{ccccccc}
 0 & 0 & 0 & 0 & 0 & 0 & 0 \\
 0 & 0 & -2 & 0 & 0 & 0 & 0 \\
 0 & 2 & 0 & 0 & 0 & 0 & 0 \\
 0 & 0 & 0 & 0 & 1 & 0 & 0 \\
 0 & 0 & 0 & -1 & 0 & 0 & 0 \\
 0 & 0 & 0 & 0 & 0 & 0 & 1 \\
 0 & 0 & 0 & 0 & 0 & -1 & 0
\end{array}
\right)
\;.
\end{eqnarray*}}
The structure constants in this basis read
\begin{eqnarray}\nonumber
&&f_{7\; 10}^{\phantom{7\; 10}13}=
-f_{7\; 11}^{\phantom{7\; 11}12}=
f_{73}^{\phantom{73}6}=
-f_{74}^{\phantom{74}5}=1\\\nonumber&&
f_{8\; 10}^{\phantom{8\; 10}12}=
f_{8\; 11}^{\phantom{8\; 11}13}=
-f_{83}^{\phantom{83}5}=
-f_{84}^{\phantom{84}6}=
f_{9\; 10}^{\phantom{9\; 10}11}=
-f_{9\; 12}^{\phantom{9\; 12}13}=
-f_{93}^{\phantom{93}4}=
f_{95}^{\phantom{95}6}=1\\\nonumber&&
f_{10\; 1}^{\phantom{10\; 1}6}=
f_{10\; 2}^{\phantom{10\; 2}5}=
-f_{11\; 1}^{\phantom{11\; 1}5}=
f_{11\; 2}^{\phantom{11\; 2}6}=
f_{12\; 1}^{\phantom{12\; 1}4}=
f_{12\; 2}^{\phantom{12\; 2}3}=
-f_{13\; 1}^{\phantom{13\; 1}3}=
f_{13\; 2}^{\phantom{13\; 2}4}=1\\
&&f_{10\; 11}^{\phantom{10\; 11}14}=
f_{12\; 13}^{\phantom{12\; 13}14}=\sqrt{3},\quad\quad
f_{78}^{\phantom{78}9}=2\\\nonumber&&
f_{14\; 1}^{\phantom{14\; 1}2}=
f_{13}^{\phantom{13}6}=
f_{14}^{\phantom{14}5}=
-f_{23}^{\phantom{23}5}=
f_{24}^{\phantom{24}6}=2/\sqrt{3}\\\nonumber&&
f_{14\; 3}^{\phantom{14\; 3}4}=
f_{14\; 5}^{\phantom{14\; 5}6}=1/\sqrt{3}
\;.
\end{eqnarray}
A basis of $G$-invariant two-, three- and four-forms is given by
\begin{equation}
\begin{array}{lllllll}
  \omega_1&=&\frac{5}{3\pi}\Big(-e^{12}+e^{34}+e^{56}\Big)&&\tilde\omega^1&=&\frac{\pi}{5\vol}\Big(e^{1234}+e^{1256}-e^{3456}\Big)\\
  \alpha_0&=&\frac{\sqrt{3}\pi}{40\vol}\Big(e^{136}+e^{145}-e^{235}+e^{246}\Big)&&\beta^0&=&\frac{10}{\sqrt{3}\pi}\Big(e^{135}-e^{146}+e^{236}+e^{245}\Big)\; .
\end{array}
\end{equation}
where
\begin{equation}
\vol=\int_Xe^{123456}=\frac{9(2\pi)^3}{20}\;.
\end{equation}
These $G$-invariant forms satisfy the half-flat mirror relations in Section~\ref{sec:mirrorgeom} with torsion parameter $e_1=1$ and intersection number $d_{111}=-100$. The non-vanishing Betti numbers are $b_0=b_6=1$ which yields the Euler number $\chi=2$,
as we would have expected from $\Gtwo/\SU{3}\cong S^6$.

The most general $G$-invariant $\SU{3}$-structures are
\begin{align}
J&=-R^2e^{12}+R^2e^{34}+R^2e^{56}=v\,\omega_1\\
\Omega&=R^3\Big((e^{136}-e^{145}+e^{235}+e^{246})+\I\, (e^{135}+e^{146}-e^{236}+e^{245})\Big)=Z\,\alpha_0+\I\,G\,\beta^0\;.
\end{align}
with associated metrics
\begin{align}
\d s_0^2
&=
R^2\,(e^1\otimes e^1+e^2\otimes e^2)
+
R^2\,(e^3\otimes e^3+e^4\otimes e^4)
+
R^2\,(e^5\otimes e^5+e^6\otimes e^6)
\;.
\end{align}
The single coset radius $R$ is related to the half-flat mirror moduli by
\begin{align}\label{eq:moduliRG2}
v=\frac{3\pi}{5}R^2
\end{align}
and
\begin{align}
Z=\frac{40\vol}{\sqrt{3}\pi}R^3\qquad\qquad
G=\frac{\sqrt{3}\pi}{10}R^3\;.
\end{align}

\section{Bianchi identity and related computations}\label{app:bianchi}
This section gives a summary of the calculation involved in solving the Bianchi identity \eqref{eq:bianchi} for the three homogeneous spaces considered. We will first focus on the connection on the tangent bundle and the computation of $\tr R^- \wedge R^-$. Then we will present the results for $\tr F \wedge F$. Finally, we insert everything into the Bianchi identity and determine the constant $\mathcal{C}$ in the Ansatz~\eqref{eq:Ansatz:hflux} for $H$.

\subsection{$\tr\, R^-\! \wedge R^-$}\label{app:trRR}

For all three spaces we have the metric of the form
\begin{align}
\d s_0^2
&=
R_1^2\,(e^1\otimes e^1+e^2\otimes e^2)
+
R_2^2\,(e^3\otimes e^3+e^4\otimes e^4)
+
R_3^2\,(e^5\otimes e^5+e^6\otimes e^6)
\; .
\end{align}
Since we are interested in performing the calculation at the nearly K\"ahler locus we set $R\equiv R_1=R_2=R_3$ so that the metric becomes the same for all three spaces. $R^-$ is calculated from the Hull connection
\begin{align}
\omega_{\phantom{-}b}^{-\phantom{b}a}= \omega_{b}^{\phantom{b}a} - \frac{1}{2}H_{cb}^{\phantom{cb}a}e^c
\;.
\end{align}
where $H=\frac{1}{3!}H_{abc}\,e^{abc}$ and $H_{cb}^{\phantom{cb}a}=H_{cbd}g^{da}$. Furthermore, $\omega$ is the Levi-Civita connection given by
\begin{align}\label{def:levicevita:appendix}
\omega_{b}^{\phantom{b}a}=\frac{1}{2} f_{cb}^{\phantom{cb}a}e^c+f_{ib}^{\phantom{ib}a}\varepsilon^i\;.
\end{align}
Here, the $\varepsilon^i$ are the coset descendants of the left-invariant Maurer-Cartan forms on $G$ \emph{in the direction of $H$}. On $G/H$ they can be expressed in terms of the basis forms $e^a$. However, we will not need these relations explicitly since the $\varepsilon^i$ will drop out of the expression for $\tr R^- \wedge R^-$.

The curvature two-form $R^{-}$ is given by
\begin{align}
(R^-)_b^{\phantom{b}a}=(\d \omega^-)_b^{\phantom{b}a} - (\omega^-)_b^{\phantom{b}c} \wedge (\omega^-)_c^{\phantom{b}a}\;,
\end{align}
where the uncommon minus sign stems from our index convention for the connection one-form.

The $H$-flux in our solution is proportional to $\alpha_0$ for each example. Using the structure constants given in Appendix \ref{app:cosets} and the definition for $\alpha_0$, this means that we can write for each coset
\begin{align}
H_{abc}= \mathcal{C}\, f_{ab}^{\phantom{ab}d} \delta_{d c} \label{Hans}
\;.
\end{align}
with a constant ${\mathcal C}$.

Let us now state the result for each case.
\subsubsection{$\SU{3}/\U{1}^2$}
Here, the Ansatz~\eqref{Hans} for the $H$-flux takes the form
\begin{align}
H = \mathcal{C}(R, \alpha')\,\frac{\vol}{\pi}\,\alpha_0
\end{align}
Evaluating $\tr\, R^-\! \wedge R^-$ at the nearly K\"ahler locus gives
\begin{align}\label{eq:trrr:su3}
\tr\, R^{-}\!\wedge R^{-} = \frac{3}{4}\left(\frac{\mathcal{C}^2}{R^4}-2\frac{\mathcal{C}}{R^2}-3\right)\frac{\vol}{\pi}\,\tilde\omega^3
\;.
\end{align}
Recall that $\tilde\omega^3=\d \alpha_0$ and, hence, this lies in the trivial cohomology class of $H^4(X)$. Consequently, the first Pontryagin class of $\SU{3}/\U{1}^2$ is $p_1(TX)=0$.

\subsubsection{$\Sp{2}/\SU{2}\times \U{1}$}
Here, the Ansatz~\eqref{Hans} for the $H$-flux reads
\begin{align}
H = \mathcal{C}(R, \alpha')\,\frac{2\,\vol}{\pi}\,\alpha_0
\end{align}
and we find
\begin{align}\label{eq:trrr:sp2}
\tr\, R^{-}\!\wedge R^{-} = 
\left( 
-48\,\tilde \omega^1
-
\left(
6\frac{\mathcal{C}^2}{R^4}
-
12 \frac{\mathcal{C}}{R^2}
-
10
\right)
\,\tilde\omega^2
\right)
\frac{\vol}{\pi}
\;.
\end{align}
Unlike for $\SU{3}/\U{1}^2$, this this represents a non-trivial cohomology class of  $H^4(X)$ given by
\begin{align}
p_1(TX)=-\frac{1}{8\,\pi^2}\left[\tr\, R^{-}\!\wedge R^{-} \right]=4\,\tilde\omega^1\; .
\end{align}

\subsubsection{$\Gtwo/\SU{3}$} 
The Ansatz~\eqref{Hans} for the $H$-flux is
\begin{align}
H = \mathcal{C}(R, \alpha')\,\frac{80\vol}{3\pi}\,\alpha_0
\;.
\end{align}
so that
\begin{align}
\tr\, R^{-}\!\wedge R^{-} = -\frac{64}{3} \left(\frac{2\mathcal{C}}{R^2}-\frac{\mathcal{C}^2}{R^4}\right)\frac{5\vol}{\pi}\,\tilde\omega^1
\;.
\end{align}
Since $\d\alpha_0=\tilde\omega^1$ the first Pontryagin class is again trivial, $p_1(TX)=0$.

\subsection{$\tr F \wedge F$}
A line bundle $L$ over the coset $G/H$ is defined by the $\dim H^2(G/H)$ integer numbers ${\bf p}=(p^r)$ such that its first Chern class is given by $c_1(L)=p^r\omega_r$. Such a line line bundle is also denoted by $L={\mathcal O}_X({\bf p})$. The curvature of a connection on $L$ is given by
\begin{align}
F=-(2\pi\I)\,p^r\omega_r\; .
\label{eq:F:app}
\end{align}
The vector bundles we construct are direct sums of line bundles
\begin{eqnarray}
V=\bigoplus_{a=1}^n {\cal O}_X({\bf p}_a)\;,
\end{eqnarray}
and are, hence, characterized by an integer matrix $(p_a^r)$. We impose that $c_1(V)\sim\sum_a{\bf p}_a=0$ so that the structure group of $V$ is $\mathrm{S}(\U{1}^n)$. 
Using the mirror half-flat geometric structure, we can express $\tr F \wedge F$ in terms of the intersection numbers by
\begin{align}\label{eq:fwedgef}
\tr F \wedge F=- 4\pi^2\, d_{rst}\sum_{a=1}^np_a^sp_a^t\,\tilde\omega^r\;.
\end{align}

\subsubsection{$\SU{3}/\U{1}^2$}
Here, there are two integers defining every line bundle and, for ease of notation we write $(p_a\,,\,q_a)=(p^1_a\,,\,p^2_a)$. The intersection numbers are given in eq.\ \eqref{eq:intersectionnums} and from direct evaluation of \eqref{eq:fwedgef} we find
\begin{equation}
\tr\, F \wedge F =- \frac{\vol}{8\pi}\left[
\sum_a(6 p_a^2 + q_a^2+6p_a q_a)\tilde{\omega}^1+
\sum_ap_a(3 p_a +2 q_a)\tilde{\omega}^2+
\frac{4}{3}\sum_a(3p_a^2 + q_a^2 + 3p_a q_a)\tilde{\omega}^3\right]
\end{equation}
This means that the second chern class of the bundle is
\begin{equation}\label{eq:trff:su3:coh}
\operatorname{ch}_2(V)=-\frac{1}{8\pi^2}\tr F \wedge F =
\frac{1}{2}\left[\sum_a(6 p_a^2 + q_a^2+6 p_a q_a)\tilde{\omega}^1+
\sum_ap_a(3 p_a +2 q_a)\tilde{\omega}^2\right]
\end{equation}
Note, in general, this represents a different cohomology class than $\tr R^-\wedge R^-$ so that solving the Bianchi identity imposes restrictions on the bundle integers $(p_a,q_a)$.

\subsubsection{$\Sp{2}/\SU{2}\times \U{1}$}
Here, a line bundle is defined by a single integer and we write $p_a=p_a^1$. Inserting the intersection numbers from \eqref{eq:intersectionnumssp2} into eq.\ \eqref{eq:fwedgef} we obtain
\begin{equation}\label{eq:trff:sp2}
\tr F \wedge F = -\frac{\vol}{\pi}\sum\limits_a p_a^2(6 \,\tilde{\omega}^1+\tilde{\omega}^2)
\end{equation}
Hence, the second Chern class of the bundle is given by
\begin{equation}\label{eq:trff:sp2:coh}
\operatorname{ch}_2(V)=-\frac{1}{8\pi^2}\tr F \wedge F = \frac{1}{2}\sum\limits_a p_a^2\;\tilde{\omega}^1\;.
\end{equation}

\subsubsection{$\Gtwo/\SU{3}$} 
In this case, the second Betti number is zero and, hence, there are no non-trivial line bundles on this coset space. However, it is still possible to solve the Bianchi identity with non-Abelian gauge bundles. An obvious choice is the (quasi) standard embedding as described in Ref.~\cite{Klaput:2011mz}. This choice has already been studied in the early work \cite{Lust:1986ix} where it was realised that the Dirac index of such bundles is
\begin{align}
\text{ind}(V_{\text{standard}})=\frac{1}{2}\chi=1\;,
\end{align}
implying one chiral family only. Another possible choice is the natural $G$-invariant connection~\cite{Klaput:2011mz} which yields a rank three bundle and solves the Hermitean Yang-Mills equations. However, this vector bundle has a Dirac index of zero and no chiral families are possible.

\subsection{Solving the Bianchi identity}\label{app:solveBianchi}
We now combine the previous results to solve the Bianchi identity
\begin{align}
\d H=\frac{\alpha'}{4}(\tr F\wedge F-\tr R^{-}\wedge R^{-})\;.
\end{align}
We will omit the case $\Gtwo/\SU{3}$ which is of no phenomenological interest in the context of our bundle construction as we have pointed out in the previous section.

We solve the Bianchi identity in three steps. Firstly, the Hermitean Yang-Mills equations are solved for the nearly K\"ahler locus $R\equiv R_i$, $\forall\; i$ and we will focus on this case. Secondly, since $\d H$ is exact $\tr\, R^-\! \wedge R^-$ and $\tr F\wedge F$ have to lie in the same cohomology class. This yield restrictions on the line bundle integers which involve the observable line bundle sum, $V=\bigoplus_{a=1}^n{\mathcal O}_X({\bf p}_a)$ and the hidden line bundle sum, $\tilde{V}=\bigoplus_{a=1}^{\tilde{n}}{\mathcal O}_X(\tilde{\bf p}_a)$. Thirdly, using these restrictions, we compute both sides of the Bianchi identity and determine the unknown constant $\mathcal{C}$ in the Ansatz~\eqref{Hans} for $H$.

\subsubsection{$\SU{3}/\U{1}^2$}
First, we note that the cohomology class of $\tr\, R^-\! \wedge R^-$ in Eq.~\eqref{eq:trrr:su3} is trivial. This means that the class of $\tr F\wedge F$ in Eq.~\eqref{eq:trff:su3:coh} needs to be trivial as well which leads to the conditions
\begin{align}\label{eq:anomalysu3cohom}
\sum\limits_{a=1}^{n}(6p_a^2 + q_a^2+6p_a q_a)+\sum\limits_{a=1}^{\tilde{n}}(6\tilde p_a^2 + \tilde q_a^2+6\tilde p_a \tilde q_a)&=0
\\
\sum\limits_{a=1}^{n}p_a(3p_a +2q_a)+\sum\limits_{a=1}^{\tilde{n}}\tilde{p}_a(3\tilde p_a +2\tilde q_a)&=0\;.
\end{align}
Together with the Ansatz~\eqref{Hans} for the flux $H$, the Bianchi identity reduces to  a quadratic equation
\begin{align}
\mathcal{C} = \frac{\alpha'}{4}\left(\mathcal{A} +\frac{3}{4}\left(-\frac{\mathcal{C}^2}{R^4}+2\frac{\mathcal{C}}{R^2}+3\right)\right)
\;.
\end{align}
for $\mathcal{C}$ where
\begin{equation}\label{eq:app:Aparameter}
\mathcal{A}(\mathbf{p},\mathbf{q}, \tilde{\mathbf{ p}}, \mathbf{\tilde q})=
-\frac{1}{6}\left[\sum\limits_{a=1}^{n}(3p_a^2 + q_a^2 + 3p_a q_a)+
\sum\limits_{a=1}^{\tilde{n}}(3\tilde p_a^2 + \tilde q_a^2 + 3\tilde p_a \tilde q_a)\right]\;.
\end{equation}
Its positive solution is
\begin{align}
\mathcal{C}(R, \alpha') = \frac{8R^4}{3\alpha'} \left(-1 + \frac{3}{8}\frac{\alpha'}{R^2} + \sqrt{1 - \frac{3}{4} \frac{\alpha'}{R^2} + \frac{3\mathcal{A}+9}{16}\, \frac{\alpha'^2}{R^4}}\right)
\;.
\end{align}
In the large radius limit, $\alpha'/R^2\ll 1$, this solution can be expanded as
\begin{align}\label{eq:squarerootexpansion:appendix}
\mathcal{C}(R, \alpha') = 
\left[\mathcal{B} 
+
\frac{(27+12\mathcal{A})}{128}\,\frac{\alpha'}{R^2}
-
\frac{(-27+24\mathcal{A}+16\mathcal{A}^2)}{4096}\frac{\alpha'^2}{R^4}
+
\mathcal{O}\left(\frac{\alpha'^3}{R^{6}}\right)\right]\,\alpha'
\; ,
\end{align}
where ${\mathcal B}=(4\mathcal{A}+9)/16$. From this we obtain the $H$-flux relevant for the four-dimensional theory is
\begin{equation}
H=\mu\alpha_0\;\; \mbox{where}\;\; \mu=\frac{\vol\alpha'}{\pi}{\mathcal B}\; .
\end{equation}

\subsubsection{$\Sp{2}/\SU{2}\times \U{1}$}
Comparing the cohomology classes of $\tr\, R^-\! \wedge R^-$ from Eq.~\eqref{eq:trrr:sp2} and $\tr F\wedge F$ from Eq.~\eqref{eq:trff:sp2:coh}, we see that equality implies
\begin{equation}
\sum\limits_{a=1}^{n}p_a^2+\sum\limits_{a=1}^{\tilde n}\tilde p_a ^2=8\; ,
\end{equation}
Together with the Ansatz~\eqref{Hans} for $H$ the Bianchi identity reduces to the quadratic equation
\begin{align}
\mathcal{C} = \frac{\alpha'}{8}\left(\mathcal{A} +10+12\frac{\mathcal{C}}{R^2}-6\frac{\mathcal{C}^2}{R^4}\right)
\end{align}
 for $\mathcal{C}$ where
\begin{align}\label{eq:app:Aparameter}
\mathcal{A}(\mathbf{p}, \tilde{\mathbf{ p}})=-\sum\limits_{a=1}^{n}p_a^2-\sum\limits_{a=1}^{\tilde n}\tilde p_a^2=-8\;.
\end{align}
Even though $\mathcal{A}$ is only a constant, we will still include it in the expressions below in order to have a notation similar to the rest of the paper.

Its positive solution is
\begin{equation}
\mathcal{C}(R, \alpha') =\frac{2R^4}{3\alpha'} \left(-1  + \frac{3}{2}\frac{\alpha'}{R^2} + \sqrt{1 - \frac{3\alpha'}{R^2} + \frac{3\, \alpha'^2}{ R^4}}\right)\; .
\end{equation}
This can once more be expanded in the large radius limit, $\alpha'/R^{2}\ll 1$, which gives
\begin{equation}
\mathcal{C}(R, \alpha') =\Biggl[\frac{1}{2}\mathcal{B} +
\frac{3}{2}\left(\frac{\mathcal{A}+10}{8}\right)\,\left(\frac{\alpha'}{R^2}\right) +
\mathcal{O}\left(\frac{\alpha'^2}{R^{4}}\right)\Biggr]\,\alpha'\; ,
\end{equation}
where $\mathcal{A}=-8$ and $\mathcal{B}=(\mathcal{A}+10)/4=1/2$. From this we obtain the $H$-flux as
\begin{equation}\label{eq:mufixed}
H=\mu\alpha_0\;\;\mbox{where}\;\;\mu=\frac{\vol\alpha'}{\pi}{\mathcal B}\; .
\end{equation}

\section{Index of the Dirac operator}\label{app:diracindex}
The chiral asymmetry of the effective four-dimensional theory is given by the index of the Dirac operator and it is, therefore, expected to count the net number of families. Hence, its knowledge is an important phenomenological constraint. In this section we will derive an expression for the index for a sum of line bundles.

On a six dimensional manifold the index is given by\cite{Nash:1991pb}
\begin{eqnarray}
\textrm{ind}(V)=-\int_X\Big(c_3(V)-\frac{1}{24}p_1(TX)c_1(V)\Big).
\end{eqnarray}
Recently, a derivation using path integral methods and the Witten index has been given in \cite{Kimura:2007xa}.

Let us first apply this to a single line bundle $L=\mathcal{O}_X({\bf p})$ with first Chern class $c_1(L)=p^r\omega_r$, where $\{\omega_r\}$ is a basis of the second cohomology. We write the first Pontryagin class as $p_1(TX)=p_{1r}(TX)\tilde\omega^r$, in terms of a basis $\{\tilde\omega^r\}$ of the fourth cohomology, dual to $\{\omega_r\}$. For line bundles we have $c_3(L)=\frac{1}{6}c_1(L)^3$, so that
\begin{align}
\textrm{ind}(L)=-\frac{1}{6}d_{rst}p^rp^sp^t+\frac{1}{24}p_{1r}p^r\; ,
\end{align}
where the intersection numbers are defined in Eq.~\eqref{eq:intersectionnumbers}. We are interested in sums of line bundles $V=\bigoplus_a\mathcal{O}_X({\bf p}_a)$ with vanishing total first Chern class, $c_1(V)\sim\sum_a{\bf p}_a=0$. For such bundles the index simplifies to
\begin{eqnarray}
\textrm{ind}(V)=-\frac{1}{6}d_{rst}\sum_ap_a^rp_a^sp_a^t\; .
\end{eqnarray}

\providecommand{\href}[2]{#2}\begingroup\raggedright\endgroup

\end{document}